\documentclass[aps,prd,twocolumn]{revtex4-1}
\usepackage{amssymb}
\usepackage{amsmath}
\usepackage{amsfonts}
\usepackage{color}
\usepackage[usenames,dvipsnames]{xcolor}
\usepackage{float}
\usepackage{accents}
\usepackage{graphicx}
\usepackage{epstopdf}
\usepackage{bm}
\usepackage[colorlinks=true,pdfstartview=FitV,linkcolor=blue,citecolor=blue,urlcolor=blue,breaklinks=true]{hyperref}

\begin{document}

\title{Self-dual effective compact and true compacton configurations in generalized Abelian Higgs models}
\author{Rodolfo Casana}
\email{rodolfo.casana@gmail.com}
\affiliation{Departamento de F\'\i sica, Universidade Federal
do Maranh\~ao, Campus Universit\'{a}rio do Bacanga, 65080-805, S\~ao Lu\'is,
Maranh\~ao, Brazil.}
\author{G. Lazar}
\email{gzsabito@gmail.com}
\affiliation{Coordena\c{c}\~ao do curso de Licenciatura em
Ci\^encias Naturais -- Campus III, Universidade Federal do Maranh\~ao,
65700-000, Bacabal, Maranh\~ao, Brazil.}
\author{Lucas Sourrouille}
\email{lsourrouille@yahoo.es}
\affiliation{IMDEA Nanociencia, Calle de Faraday, 9, Cantoblanco,
28049, Madrid, Spain}
\affiliation{IFISUR, Departamento de F\'\i sica (UNS-CONICET), Avenida Alem 1253, Bah\'\i a Blanca, Buenos Aires, Argentina. }

\begin{abstract}
We have studied the existence of self-dual effective compact and  true
compacton configurations in Abelian Higgs models with generalized dynamics.
We have named of an  effective compact solution the one whose profile
behavior is very similar to the one of a compacton structure but still
preserves a  tail in its asymptotic decay. In particular we have investigate
the electrically neutral configurations of the Maxwell-Higgs and
Born-Infeld-Higgs models and the electrically charged ones of the
Chern-Simons-Higgs and Maxwell-Chern-Simons-Higgs models. The generalization
of the kinetic terms is performed by means of dielectric functions in gauge
and Higgs sectors. The implementation of the BPS formalism without the need
to use a specific \textit{Ansatz} has leaded us to the explicit determination
of the dielectric function associated to the Higgs sector to be proportional
to $\lambda |\phi |^{2\lambda -2}$, $\lambda >1$. Consequently, the followed
procedure allows us to determine explicitly new families of self-dual
potentials for every model. We have also observed that for sufficiently large
values of $\lambda$ every model supports effective compact vortices. The true
compacton solutions arising for $\lambda= \infty $ are analytical.
Therefore, this new self-dual structures enhance the space of BPS solutions
of the Abelian Higgs models and they probably will imply in interesting
applications in physics and mathematics.
\end{abstract}

\maketitle

\section{Introduction}

Topological defects produced by field theories including generalized kinematic
terms have been an issue of great interest in the latest years. Usually these
models include higher derivatives dynamic terms, but sometimes the
generalizations is caused by the introduction of some generalized parameter or
functional in  the kinetic terms. These modified theories named as
\textit{k}-theories arose initially as effective cosmological models for
inflationary evolution \cite{Kinflation}. Later the \textit{k}-theories were
permeating other issues of field theory and cosmology such as: dark matter
\cite{DarkMatter}, strong gravitational waves \cite{Gwaves}, the tachyon matter
problem \cite{12} and ghost condensates \cite{13}.  It is worth emphasizing  the possibility that these theories can arise naturally within the context of string  theory. Several studies concerning its topological structure have shown \textit{k}-theories support topological soliton both in models of matter as in gauged models \cite{BAi,SG,SG1}, in general, they can present some new characteristics when compared those of the usual ones \cite{B}.

Compactons  were defined as solitons with finite wavelength in the pioneer work
\cite{AM} and so far they have been the subject of several studies, since
models containing topological defects have been used to represent particles and
cosmological objects such as cosmic strings \cite{NO}. A particular arrangement
of particles can be represented with a group of compactons and in this case we
will not have the problem of the superposition of particles (or defects) due
that compactons do not carry a ``tail" in its asymptotic decay. We also point
out that compact vortices and skyrmions are intrinsically connected with recent
advances in the miniaturization of magnetic materials at the nanometric
scale for spintronic applications \cite{Jub,Fert,Romming}.  Compact
topological defects have gained greater attention as effective low-energy
models for QCD concerning skyrmions, where non-perturbative results at the
classic level have been reached \cite{Sanchez-Guillen}. Compact solutions were
also successfully employed in the description of boson stars \cite{Boson stars}
and in baby Skyrme models \cite{BSKY1,BSKY2}.

At the classical level there is a widely employed mechanism to achieve field
equations, namely the Bogomol'nyi-Prasad-Sommerfield (BPS) formalism
\cite{BPS}. The BPS method consists in building a set of first-order
differential equations which solve as well the second-order Euler-Lagrange
equations. One interesting aspect of this mechanism is that all the equations
are build up for static field configurations. As a consequence, the first-order
equations of motion coming out from the BPS formalism describe field
configurations minimizing the total system energy. The static characteristic of
the fields in the BPS limit have been applied to investigate topological
defects in several frameworks. For example, in the context of planar gauge
theories, vortices structures arise from the BPS equations, specially, magnetic
vortices were found in Maxwell-Higgs electrodynamics \cite{NO}. Also we can
mention the  Chern-Simons-Higgs electrodynamics \cite{cshv} and   \cite{mcshv}
the Maxwell-Chern-Simons-Higgs model both describing electrically charged
magnetic vortices. Other interesting framework involving first-order BPS
solutions are the nonlinear sigma models (NL$\sigma$M) \cite{polyakov} in the
presence of a gauge field. These theories have been widely applied in the study
of field theory and condensed matter physics \cite{CMP}. We can mention in this
sense, the topological defects in a $O(3)$ nonlinear sigma model with the
Maxwell term, as it is shown in \cite{schroers,mukherjee2}. Concerning the
Chern-Simons term, topological and nontopological defects were analyzed in
\cite{ghosh,mukherjee1}. As well the gauged $O(3)$ sigma model with both,
Maxwell and the Chern-Simons terms was studied in Refs.
\cite{sigmaMCSH1,sigmaMCSH2}.

The existence of vortex solutions with compact-like profiles in $k$%
-generalized Abelian Maxwell-Higgs model were studied in Ref. \cite{D1,D1x}
and in $k$-generalized Born-Infeld model \cite{D2}, however, only in Ref.
\cite{D1x} was found first-order vortices with compact-like profiles.
Therefore, the aim of this manuscript is to study the topological vortices
engendered by the self-dual configurations obtained from the generalization
of the following Abelian Higgs models: Maxwell-Higgs (MH), Born-Infeld-Higgs
(BIH), Chern-Simons-Higgs (CSH) and Maxwell-Chern-Simons-Higgs (MCSH).
Firstly, we have performed a consistent implementation of the Bogomol'nyi
method for every model and obtained the respective generalized self-dual or BPS
equations.  The developing of the BPS formalism has allowed to fix the form of
the function $\omega (|\phi |)$ -composing the generalized term $\omega(|\phi|)
|D_{\mu}\phi |^{2}$- which only can be proportional to
$\lambda|\phi|^{2\lambda-2}$ for $\lambda >1$ (a similar result was obtained in
Ref. \cite{L2}). Secondly, we use the usual vortex \textit{Ansatz} to obtain
the self-dual equations describing axially symmetric configurations.  We have observed that independently of the model an effective compact  behavior of
the vortices  arises for a sufficiently large value of the parameter $\lambda$
and only in the limit $\lambda\rightarrow\infty$ the true compacton structures
are achieved. Finally, we give our remarks and conclusions.

\section{The Maxwell-Higgs case}

The Maxwell-Higgs model is a classical field theory where the gauge field
dynamics is controlled by the Maxwell term and the matter field is represented
by the complex scalar  Higgs field. The model presents vortex  solutions when
it is endowed with a fourth-order self-interacting potential promoting a
spontaneous symmetry  breaking. Although the model seems very simple  it
presents characteristics very similar to the phenomenological Ginzburg-Landau model for superconductivity  \cite{Abrikosov} or superfluidity in He$^{4}$. The applications of the Maxwell-Higgs model extend from the condensed matter to inflationary cosmology  \cite{Vilenkin} or as an effective field theory for cosmic strings \cite{NO}.

The generalized Maxwell-Higgs model \cite{Bazeia1} is described by the
following Lagrangian density%
\begin{equation}
\mathcal{L}=-\frac{G(|\phi |)}{4}F_{\mu \nu }F^{\mu \nu }+\omega (|\phi
|)|D_{\mu }\phi |^{2}-V(|\phi |)~.  \label{Acg1a}
\end{equation}%
The nonstandard dynamics is introduced by two non-negative functions
$G(|\phi|)$ and $\omega(|\phi|)$ depending of the Higgs field.  The Greek index
running from $0$ to $2$. The vector $A_{\mu}$ is the electromagnetic field,
the Maxwell strength tensor is $F_{\mu\nu}=\partial _{\mu}A_{\nu }-
\partial _{\nu }A_{\mu }$, and $D_{\mu}\phi$ defines the covariant derivative
of the Higgs field $\phi$,
\begin{equation}
D_{\mu }\phi =\partial _{\mu }\phi -ieA_{\mu }\phi .
\end{equation}%
The function $V(\vert\phi\vert)$ is a self-interacting scalar potential.

From (\ref{Acg1a}), the gauge field equation reads
\begin{equation}
\partial _{\nu }\left( GF^{\nu \mu }\right)=e\omega J^{\mu } ,  \label{ge0a}
\end{equation}%
where $\omega J^{\mu }$ is the conserved current density, i.e., $%
\partial_\mu(\omega J^{\mu})=0$, and $J^{\mu}$ is the usual current density
\begin{equation}
J^{\mu }=i[\phi (D^{\mu }\phi )^{\ast }-\phi ^{\ast }(D^{\mu }\phi )].
\end{equation}

Along the remain of the section, we are interested in time-independent
soliton solutions that ensure the finiteness of the action engendered by (%
\ref{Acg1a}). Then, from Eq. (\ref{ge0a}), we read the static Gauss law
\begin{equation}
\partial _{k}\left( G\partial _{k}A_{0}\right) =2e^{2}\omega A_{0}\left\vert
\phi \right\vert ^{2},  \label{geEq1a}
\end{equation}%
and the respective Amp\`{e}re law
\begin{equation}
\epsilon _{kj}\partial _{j}\left( GB\right) =e\omega J_{k}.  \label{geEq2a}
\end{equation}

It is clear from the Gauss law that $e\omega J_{0}$ stands for the electric
charge density, so that the total electric charge of the configurations is
\begin{equation}
Q=2e^2\int d^{2}x\,\omega A_{0} \vert\phi\vert ^{2}.
\end{equation}%
which is shown to be null ($Q=0$) by integration of the Gauss law under
suitable boundary conditions for the fields at infinity, i.e., $%
A_{0}\rightarrow 0$, $\phi \rightarrow cte$ and $G(|\phi|)$ a well behaved
function. Therefore, the field configurations will be electrically neutral,
like it happens in the usual Maxwell-Higgs model.

The fact the configurations being electrically neutral is compatible with
the gauge condition, $A_{0}=0,$ which satisfies identically the Gauss law (%
\ref{geEq1a}). With the choice $A_{0}=0$, the static and electrically
neutral configurations are described by the Amp\`{e}re law (\ref{geEq2a})
and the reduced equation for the Higgs field
\begin{equation}
D_{k}\left( \omega D_{k}\phi \right) -\frac{1}{2}B^{2}\frac{\partial G}{%
\partial \phi ^{\ast }}-\frac{\partial \omega }{\partial \phi ^{\ast }}%
|D_{k}\phi |^{2}-\frac{\partial V}{\partial \phi ^{\ast }}=0~.
\end{equation}

To implement the BPS formalism, we first establish the energy for the static
field configuration in the gauge $A_{0}=0$, so it reads
\begin{equation}
E= \int \!d^{2}x\!\left[ \frac{G}{2}B^{2}+\omega |D_{k}\phi |^{2}+V\right] .
\label{EJPa}
\end{equation}%
To proceed, we need the fundamental identity
\begin{equation}
|D_{i}\phi |^{2}=|D_{\pm }\phi |^{2}\pm eB|\phi |^{2}\pm \frac{1}{2}\epsilon
_{ik}\partial _{i}J_{k}~,  \label{iden}
\end{equation}%
where $D_{\pm }\phi =D_{1}\phi \pm iD_{2}\phi $. With it, the energy (\ref%
{EJPa}) is written as being
\begin{eqnarray}
E &=&\int \!d^{2}x\left[ \frac{G}{2}B^{2}+V(|\phi |)+\omega |D_{\pm }\phi
|^{2}\right.  \label{EJP2a} \\
&&~\ \ \ \ \ \ \ \ \left. \pm e\omega B\left\vert \phi \right\vert ^{2}\pm
\frac{1}{2}\omega \epsilon _{ik}\partial _{i}J_{k}\right] ~.  \notag
\end{eqnarray}

We observe the term $\omega \epsilon _{ik}\partial _{i}J_{k}$ precludes the
implementation of the BPS procedure, i.e, \ to express\ the integrand as a
sum of squared terms plus a total derivative plus a term proportional to the
magnetic field. This inconvenience already was observed in \cite{AHEP-CASANA}%
, such a problem was circumvented by analyzing only axially symmetric
solutions in polar coordinates.

The key question about the functional form of $\omega (|\phi |)$ allowing a
well defined implementation of the BPS formalism was solved in Ref. \cite{L2}%
. In the following we reproduce some details of the looking for the function
$\omega (|\phi |)$. The starting point is the following expression:
\begin{equation}
\omega \epsilon _{ik}\partial _{i}J_{k}=\epsilon _{ik}\partial _{i}(\omega
J_{k})-\epsilon _{ik}(\partial _{i}\omega )J_{k}.  \label{eqw1}
\end{equation}%
By considering $\omega $ be a explicit function of $|\phi |^{2}$, after some
algebraic manipulations, the last term $\epsilon _{ik}(\partial _{i}\omega
)J_{k}$ becomes expressed as%
\begin{equation}
\epsilon _{ik}(\partial _{i}\omega )J_{k}=\left\vert \phi \right\vert ^{2}%
\frac{\partial \omega }{\partial \left\vert \phi \right\vert ^{2}}\epsilon
_{ik}\partial _{i}J_{k}+2eB\left\vert \phi \right\vert ^{4}\frac{\partial
\omega }{\partial \left\vert \phi \right\vert ^{2}}~,  \label{eqw3}
\end{equation}%
which after substituted in Eq. (\ref{eqw1}) allows to obtain
\begin{equation}
\left( \omega +|\phi |^{2}\frac{\partial \omega }{\partial |\phi |^{2}}%
\right) \epsilon _{ik}\partial _{i}J_{k}=\epsilon _{ik}\partial _{i}(\omega
J_{k})-2eB|\phi |^{4}\frac{\partial \omega }{\partial |\phi |^{2}}~.
\label{eqw4}
\end{equation}
At this point, we establish the function $\omega $ to satisfy the following
condition:
\begin{equation}
\omega +\left\vert \phi \right\vert ^{2}\frac{\partial \omega }{\partial
\left\vert \phi \right\vert ^{2}}=\lambda \omega ,~\lambda >0,  \label{eqw5}
\end{equation}%
whose solutions provides the explicit functional form of $\omega (|\phi |)$
to be
\begin{equation}
\omega (|\phi |)=\lambda \frac{|\phi |^{2\lambda -2}}{v^{2\lambda -2}},
\label{omegaa}
\end{equation}%
it will guarantee that the vacuum expectation value of the Higgs field be $%
|\phi |=v$.

With the key condition (\ref{eqw5}), the Eq. (\ref{eqw4}) allows to express
the term $\omega \epsilon _{ik}\partial _{i}J_{k}$ in the following way
\begin{equation}
\omega \epsilon _{ik}\partial _{i}J_{k}=\frac{1}{\lambda }\epsilon
_{ik}\partial _{i}\left( \omega J_{k}\right) -2ev^{2}\left( \lambda
-1\right) \frac{|\phi |^{2\lambda }}{v^{2\lambda }}B.  \label{eqw7}
\end{equation}

By putting the expression (\ref{eqw7}) in the energy (\ref{EJP2a}), it
becomes
\begin{eqnarray}
E &=&\int \!d^{2}x\,\left[ \frac{G}{2}B^{2}+V(|\phi |)+\omega |D_{\pm }\phi
|^{2}\right.  \label{EJP3a} \\
&&\hspace{1.25cm}\left. \pm ev^{2}\frac{|\phi |^{2\lambda }}{v^{2\lambda }}%
B\pm \frac{1}{2\lambda }\epsilon _{ik}\partial _{i}\left( \omega
J_{k}\right) \right] .  \notag
\end{eqnarray}%
We now manipulate the two first terms in such a form the energy can be
written as%
\begin{eqnarray}
E &=&\int \,\,d^{2}x\left[ \frac{G}{2}\left( B\mp \sqrt{\frac{2V}{G}}\right)
^{2}+\omega |D_{\pm }\phi |^{2}\right.  \label{EPJ4a} \\
&&\left. \pm B\left( \sqrt{2GV}+ev^{2}\frac{|\phi |^{2\lambda }}{v^{2\lambda
}}\right) \pm \frac{1}{2\lambda }\epsilon _{ik}\partial _{i}\left( \omega
J_{k}\right) \right] .  \notag
\end{eqnarray}%
With the objective the integrand to have a term proportional to the magnetic
field, we impose that the factor multiplying it be a constant, i.e.,
\begin{equation}
\sqrt{2GV}+ev^{2}\frac{|\phi |^{2\lambda }}{v^{2\lambda }}=ev^{2}.
\end{equation}%
Consequently, the self-dual potential $V(|\phi |)$ becomes
\begin{equation}
V(|\phi |)=\frac{1}{G}U^{(\lambda )}(|\phi |),  \label{potsbpsa}
\end{equation}%
where we have defined the potential $U^{(\lambda )}(|\phi |)$ given by
\begin{equation}
U^{(\lambda )}(|\phi |)=\frac{e^{2}v^{4}}{2}\left( 1-\frac{|\phi |^{2\lambda
}}{v^{2\lambda }}\right) ^{2}.  \label{mhlambda}
\end{equation}%
We can note that for $\lambda =1$ it becomes the self-dual potential of the
Maxwell-Higgs model.

Hence, the energy (\ref{EPJ4}) reads%
\begin{eqnarray}
E &=& \int \!\!d^{2}x\Bigg\{\!\!\pm ev^{2}B+\omega |D_{\pm }\phi |^{2} \\
&& \hspace{1cm} +\frac{G}{2}\left( B\mp \frac{\sqrt{2U^{(\lambda )}}}{G}%
\right) ^{2}\pm \frac{1}{2\lambda }\epsilon _{ik}\partial _{i} ( \omega
J_{k} ) \!\Bigg\} .  \notag
\end{eqnarray}%
Now by imposing appropriate boundary conditions, the contribution to the
total energy of the total derivative is null and the energy has a lower
bound proportional to the magnitude of the magnetic flux,
\begin{equation}
E\geq \pm ev^{2}\!\!\int \!\!d^{2}xB=ev^{2}|\Phi |~,  \label{lower}
\end{equation}%
where for positive flux we choose the upper signal, and for negative flux we
choose the lower signal.

The lower bound is saturated by fields satisfying the first-order
Bogomol'nyi or self-dual equations \cite{BPS}
\begin{equation}
D_{\pm }\phi =0,  \label{BPS1a}
\end{equation}%
\begin{equation}
B=\pm \frac{ev^{2}}{G}\left( 1-\frac{|\phi |^{2\lambda }}{v^{2\lambda }}%
\right) .  \label{BPS2a}
\end{equation}
The function $G(|\phi |)$ must be a function providing a finite magnetic
field such that $B(|\vec{x}|\rightarrow \infty)\rightarrow 0$ sufficiently
rapid to provide a finite total magnetic flux.

In the BPS limit the energy (\ref{EJPa}) provides the energy density of the
self-dual configurations
\begin{equation}
\varepsilon _{_{BPS}}=\frac{2U^{(\lambda )}}{G}+\lambda \frac{|\phi
|^{2\lambda -2}}{v^{2\lambda -2}}\left\vert D_{k}\phi \right\vert ^{2},
\label{ebpsmh}
\end{equation}%
it will be finite and positive-definite for $\lambda >0$. We here also
require the function $G(|\phi |)$ yielding a finite BPS energy density such
that $\varepsilon_{_{BPS}}(|\vec{x}|\rightarrow \infty)\rightarrow 0$
sufficiently rapid to provide a finite total energy.

\subsection{Maxwell-Higgs effective compact vortices for $\protect\lambda$ finite}

In the following, with loss of generality, we have chosen $G(|\phi |)=1$
(for this one and for all other models analyzed along the manuscript) with
the aim to study the influence of the generalized dynamic in Higgs sector in
the formation of effective compact vortices and true compactons. Such a
generalization provided by the function $\omega(|\phi |)$ defined in Eq.
(\ref{omegaa}) and its effects in the formation of effective compact vortices
apparently remains unexplored in the literature.

Thus, we seek axially symmetric solutions according to the usual vortex
\textit{Ansatz} \cite{NO}
\begin{eqnarray}
\phi (r,\theta )= vg(r)e^{in\theta }, \quad
A_{\theta }(r)= -\frac{a(r)-n}{er}, \label{ax2}
\end{eqnarray}
with $n=\pm 1,\pm 2,\pm 3...$ standing for the winding number of the vortex
solutions.

The profiles $g(r)$ and $a(r)$ are regular functions describing
solutions possessing finite energy and obeying the boundary conditions,
\begin{eqnarray}
g(0) &=&0,\;\;a(0)=n,  \label{bcx1} \\[0.2cm]
g(\infty ) &=&1\text{\textbf{,}}\;\;a(\infty )=0.  \label{bcx2}
\end{eqnarray}%
Under the \textit{Ansatz} (\ref{ax2}), the magnetic field reads
\begin{equation}
B(r)=-\frac{1}{er}\frac{da}{dr}.  \label{B}
\end{equation}
The correspondent quantized magnetic flux is given by
\begin{equation}
\Phi =\int \!d^{2}xB=\frac{2\pi }{e}n,  \label{flujo}
\end{equation}%
as expected.

The BPS equations (\ref{BPS1a}) and (\ref{BPS2a}) are written as
\begin{eqnarray}
g^{\prime } &=&\pm \frac{ag}{r},  \label{bpsx02} \\[0.2cm]
-\frac{a^{\prime }}{r} &=&\pm e^{2}v^{2}(1-g^{2\lambda }).  \label{bpsx2}
\end{eqnarray}%
The upper (lower) signal corresponds to the vortex (antivortex) solution
with winding number $n>0$ ($n<0$).

The self-dual energy density (\ref{ebpsmh}) is expressed by
\begin{equation}
\varepsilon _{_{BPS}}=e^{2}v^{4}\left( 1-g^{2\lambda }\right) ^{2}+2\lambda
v^{2}g^{2\lambda -2}\left( \frac{ag}{r}\right) ^{2}.  \label{enn2}
\end{equation}%
It will be finite and positive-definite for $\lambda \geq 1$.

The total energy of the self-dual solutions is given by the lower bound (\ref%
{lower}),
\begin{equation}
E_{_{BPS}}=\pm ev^{2}\Phi _{B}=\pm 2\pi v^{2}n,  \label{bpsenergy}
\end{equation}%
it is proportional to the winding number of the vortex solution, as expected.

The behavior of $g(r)$ and $a(r)$ near the boundaries can be easily
determined by solving the self-dual equations (\ref{bpsx02}) and (\ref{bpsx2}%
) around the boundary conditions (\ref{bcx1}) and (\ref{bcx2}). Then, for $%
r\rightarrow 0$, the profiles behave as
\begin{eqnarray}
g(r) &\approx &C_{n}r^{n}+...  \label{Assym_g_MH} \\[0.06in]
a(r) &\approx &n-\frac{e^{2}v^{2}}{2}r^{2}+...  \label{Assym_a_MH}
\end{eqnarray}%
where the constant  $C_{n}>0$ is computed numerically.

\begin{figure}[]
\centering\includegraphics[width=8.6cm]{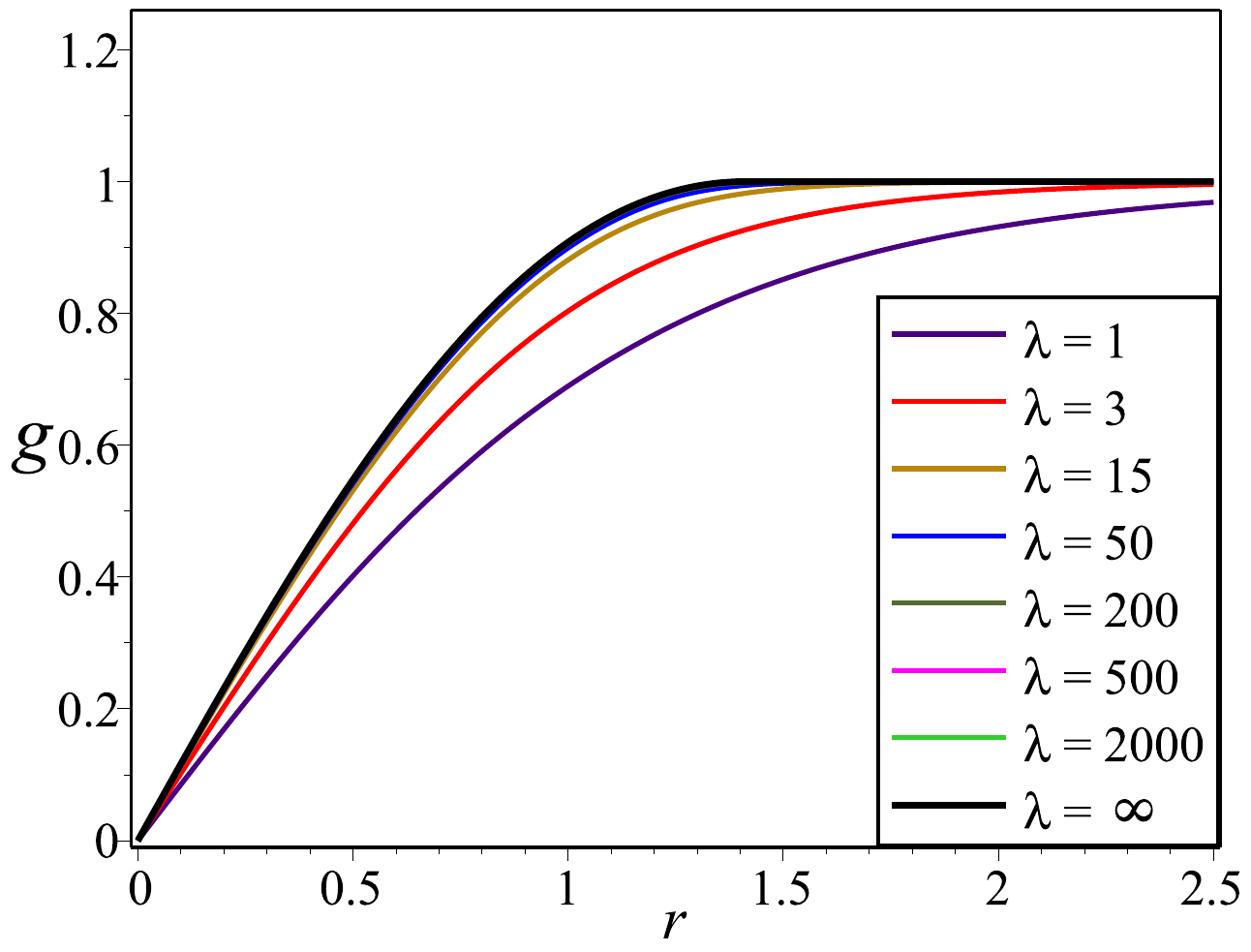}\vspace{0.15cm} %
\centering\includegraphics[width=8.6cm]{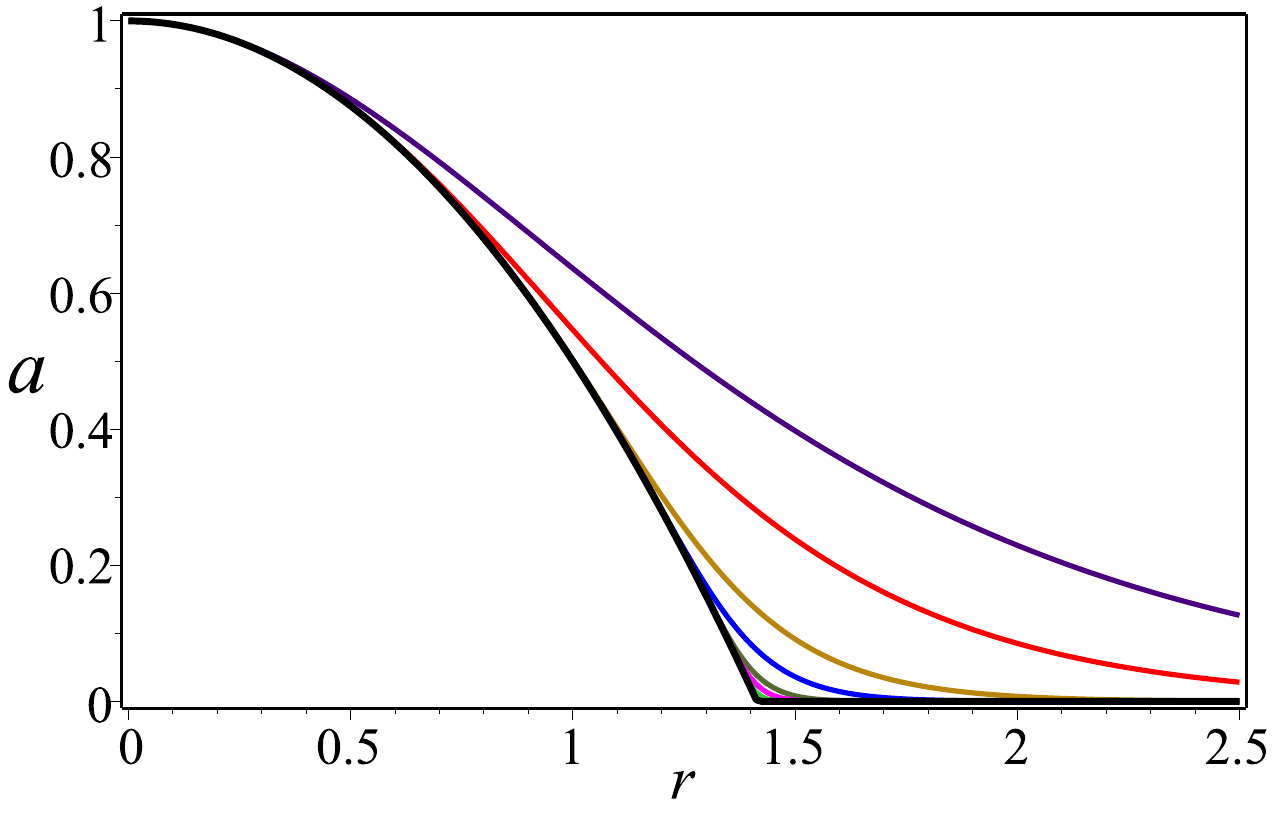}
\caption{The profiles $g(r)$ (upper) and $a(r)$ (lower) coming from
generalized Maxwell-Higgs model (\protect\ref{Acg1a}) with $G(g)=1$ and $\protect\omega(g)=\protect\lambda g^{2\protect\lambda-2}$. Observe that $\lambda=1$ (indigo lines) represents the usual MH model and the true compacton solution is given by $\lambda=\infty$ (black lines).}
\label{fig01}
\end{figure}

\begin{figure}[]
\centering\includegraphics[width=8.6cm]{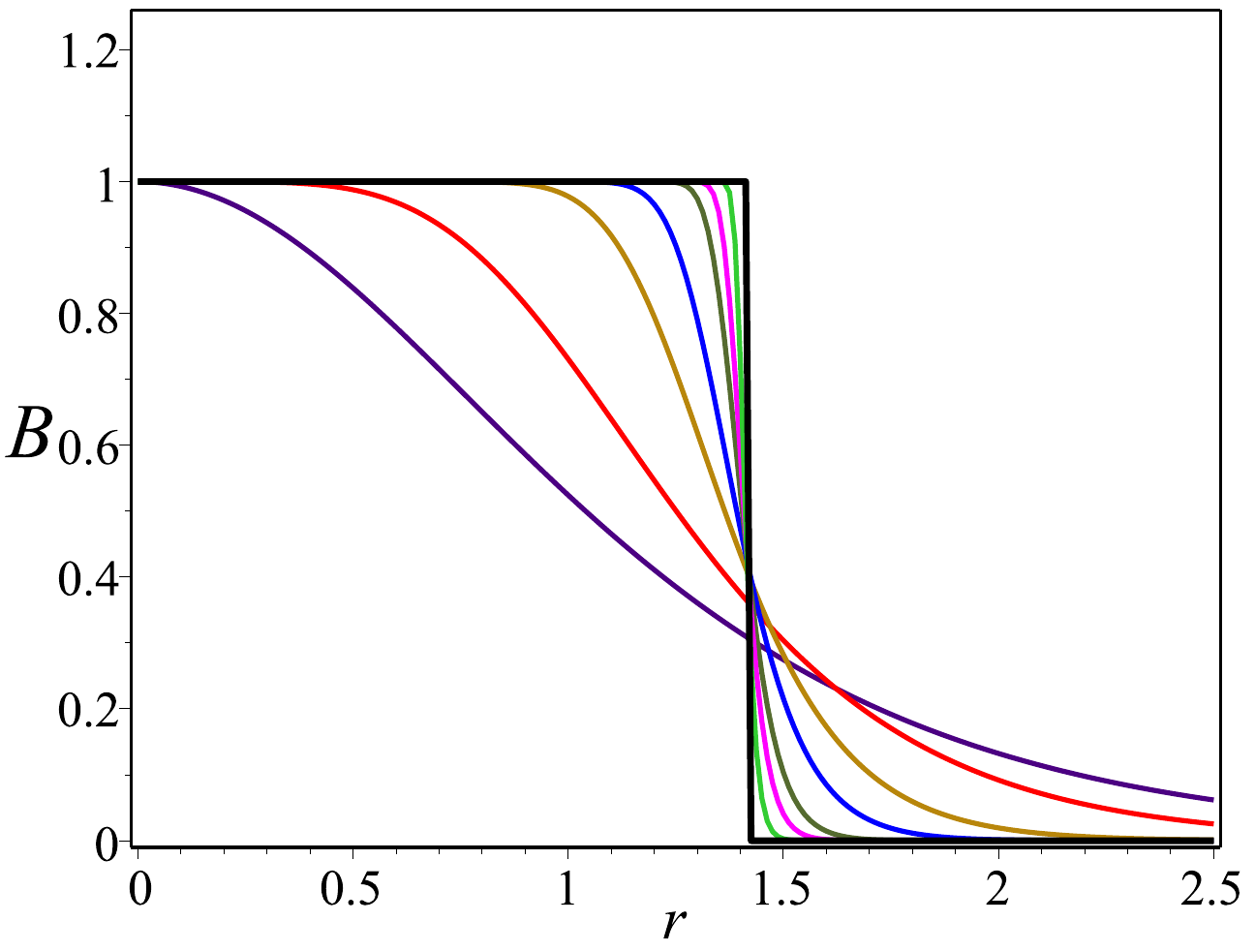}\vspace{0.23cm} %
\centering\includegraphics[width=8.6cm]{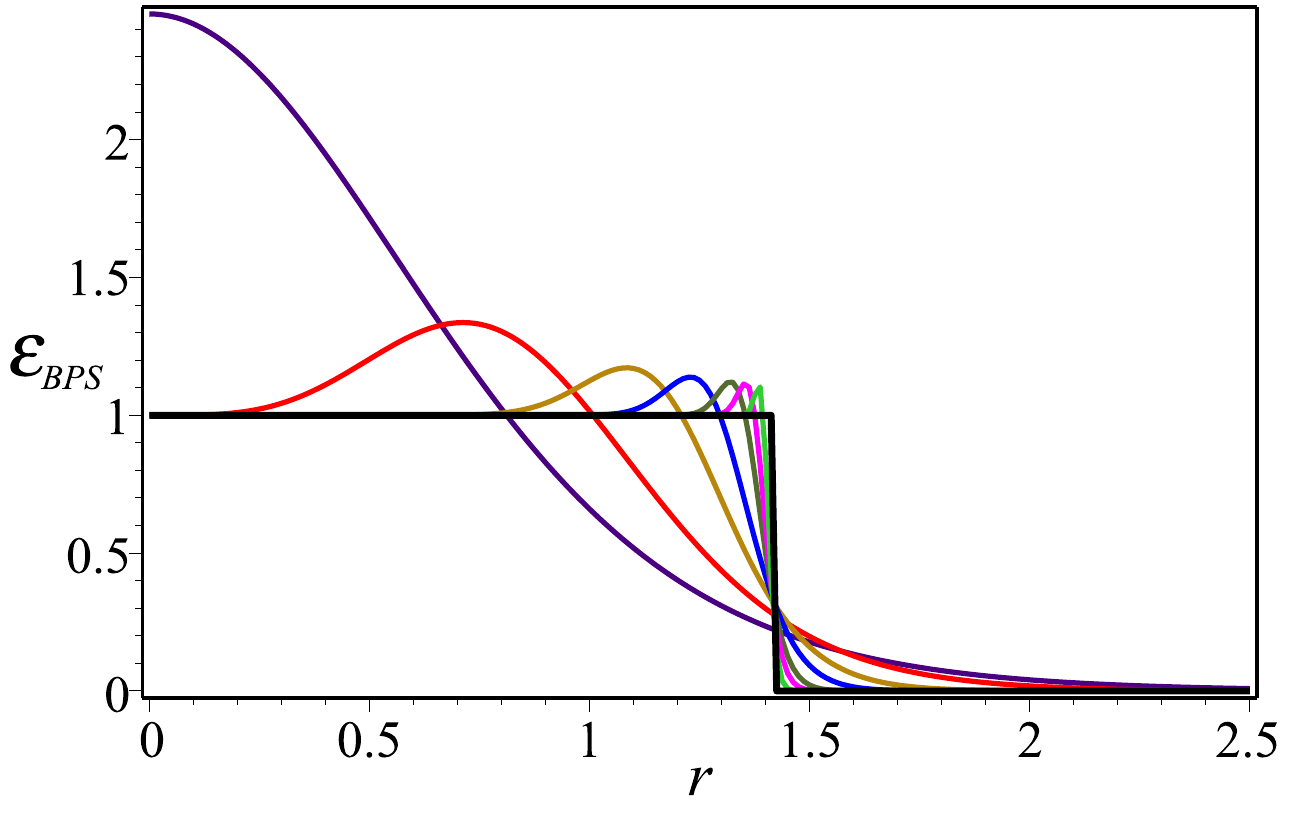}
\caption{The magnetic field $B(r)$ (upper) and the BPS energy density $\protect\varepsilon _{_{BPS}}(r)$ (lower) coming from generalized
Maxwell-Higgs model (\protect\ref{Acg1a}) with $G(g)=1$ and $\protect\omega (g)=\protect\lambda g^{2\protect\lambda -2}$. Observe that $\lambda=1$ (indigo lines) represents the usual MH model and the true compacton solution is given by $\lambda=\infty$ (black lines).}
\label{fig02}
\end{figure}

On the other hand, when $r\rightarrow \infty $, they behave as
\begin{eqnarray}
g(r) &\approx &1-\frac{C_{\infty }}{\sqrt{r}}e^{-mr}, \\[0.15cm]
a(r) &\approx &mC_{\infty }\sqrt{r}e^{-mr},
\end{eqnarray}%
the constant $C_{\infty}$ is determined numerically and $m$, the self-dual
mass, is given by
\begin{equation}
m=ev\sqrt{2\lambda },  \label{massMH}
\end{equation}%
remembering that $ev\sqrt{2}$ is the mass scale of the usual Maxwell-Higgs
model.  The influence of the generalization in the mass scale
explains the changes in the vortex-core size for large values of $\lambda$
observed in the Figs. \ref{fig01} and \ref{fig02}.

\subsection{Maxwell-Higgs compactons for $\protect\lambda=\infty$}

By considering the profile $0\leq g(r)<1$, in the limit $\lambda
\rightarrow\infty $ the potential (\ref{mhlambda}) acquires the following
form%
\begin{equation}
U^{(\infty)}(g)=\frac{e^{2}v^{4}}{2}\Theta( 1-g) ,
\end{equation}%
where $\Theta( 1-g)$ is the Heaviside function.

The BPS equations (\ref{bpsx02}) and (\ref{bpsx2}), in the limit $\lambda
\rightarrow\infty $ are written as
\begin{eqnarray}
g^{\prime }&=&\pm \frac{ag}{r},  \label{cp1} \\[0.2cm]
-\frac{a^{\prime }}{r}&=&\pm e^{2}v^{2}\Theta( 1-g).  \label{cp2}
\end{eqnarray}

The boundary conditions for compacton solutions are
\begin{eqnarray}
g(0) &=&0,\;\;a(0)=n,  \label{bcx1c} \\[0.25cm]
g(r) &=&1,\;\;a(r)=0,\;r_{c}\leq r<\infty .  \label{bcx2c}
\end{eqnarray}%
The radial distance $r_{c}<\infty $ is the value where the profile $g(r)$
reaches the vacuum value and the gauge field profile $a(r)$ becomes null.

The solutions (for $n>0$) of the compacton BPS equations (\ref{cp1}) and (%
\ref{cp2}) provides analytical profiles for the Higgs and gauge field,
\begin{eqnarray}
g^{(\infty )}(r)&=&\left(\frac{r}{r_{c}}\right)^n \exp \left[ \frac{n}{2}\left( 1-%
\frac{r^{2}}{r_{c}^{2}}\right) \right] \Theta (r_{c}-r) \\[0.2cm]
& & +\Theta (r-r_{c}),\nonumber\\[0.2cm]
a^{(\infty )}(r)&=&n\left( 1-\frac{r^{2}}{r_{c}^{2}}\right) \Theta ({r_{c}}-{r}),
\end{eqnarray}
where the radial distance $r_{c}$ is given by
\begin{equation}
r_{c}=\frac{\sqrt{2n}}{\left\vert ev\right\vert }.
\end{equation}%
The magnetic field and BPS energy density of the Maxwell-Higgs compacton are
\begin{eqnarray}
B^{(\infty )}(r) &=&ev^{2}\Theta (r_{c}-r), \\[0.25cm]
\varepsilon _{_{BPS}}^{(\infty )}(r) &=&e^{2}v^{4}\Theta (r_{c}-r).
\end{eqnarray}

The numerical solutions (for all model analyzed in the manuscript) were performed using the routines for boundary value problems of the software Maple 2015. We have chosen the upper signals in BPS equations (\ref{bpsx02}) and (\ref{bpsx2}). We have fixed $e=v=1$, the winding number $n=1$ and calculated the numerical solutions for some finite values of $\lambda$. The profiles for the Higgs and gauge fields are given in Fig. \ref{fig01} and, the correspondent ones for the magnetic field and the self-dual energy density are depicted in Fig. \ref{fig02}.

The numerical analysis shows that for sufficiently large but finite values of
$\lambda$, the profiles  are very alike to compacton solution ones, we have
named them Maxwell-Higgs effective compact vortices. The true Maxwell-Higgs
compacton is formed when $\lambda=\infty$ (see black line profiles in Figs.
\ref{fig01} and \ref{fig02}).

\section{The Born-Infeld-Higgs case}

The Born-Infeld theory is a nonlinear electrodynamic that was introduced to
remove the divergence of the electron self-energy \cite{BI}. It is the only
completely exceptional nonlinear electrodynamics because to the absence of
shock waves and birefringence in its propagation properties \cite{Boillat}.
Concerning topological defects in the Born-Infeld-Higgs model, vortex
solutions were found in \cite{BI-VORTEX}. One generalization of BIH model
was firstly done in \cite{D2} but no self-dual solutions were found. On the
other hand, the self-dual or BPS topological vortex solutions were found in
a generalized Born-Infeld-Higgs model introduced in Ref. \cite{Casana}.

The Lagrangian density of our ($2+1$)-dimensional theory is written as
\begin{equation}
\mathcal{L}=\beta ^{2}\left( 1-\mathcal{R}\right) +\omega \left( \left\vert
\phi \right\vert \right) \left\vert D_{\mu }\phi \right\vert ^{2}-W\left(
\left\vert \phi \right\vert \right) \text{,}  \label{eq:action}
\end{equation}%
where we go to consider the function $\omega \left( \left\vert \phi
\right\vert \right) $ given by Eq. (\ref{omegaa}). We have also defined the
following functions
\begin{eqnarray}
\mathcal{R} &=&\sqrt{1+\frac{G\left( \left\vert \phi \right\vert \right) }{%
2\beta ^{2}}F_{\mu \nu }F^{\mu \nu }}, \\
W\left( \left\vert \phi \right\vert \right) &=&\beta ^{2}\left[ 1-V\left(
\left\vert \phi \right\vert \right) \right] .
\end{eqnarray}%
The generalized potential $W\left( \left\vert \phi \right\vert \right) $, a
nonnegative function, inherits its structure from the function $V(|\phi |)$,
which is restricted by the condition $0<V(|\phi |)\leq 1$, so $W(\phi )>0$.
The Born-Infeld parameter $\beta $, provides modified dynamics for both
scalar and gauge fields further enriching the family of possible models.

From the action (\ref{eq:action}) the gauge field equation of motion reads
\begin{equation}
\partial _{\nu }\left( \frac{G}{\mathcal{R}}F^{\nu \mu }\right) =e\omega
J^{\mu }.  \label{eq:gauge}
\end{equation}%
We are interested in stationary solutions, so the Gauss law becomes
\begin{equation}
\partial _{j}\left( \frac{G}{\mathcal{R}}\partial _{j}A_{0}\right) =2\omega
e^{2}A_{0}\left\vert \phi \right\vert ^{2}.  \label{eq:Gauss}
\end{equation}%
Similarly to it happens in Maxwell-Higgs model, the field configurations are
electrically neutral therefore we go to work in the gauge $A_{0}=0$.

Consequently, at static regime, in the gauge $A_{0}=0$, the Amp\`{e}re law
is given by%
\begin{equation}
\epsilon _{kj}\partial _{j}\left( \frac{G}{\mathcal{R}}B\right) -e\omega
J_{k}=0,  \label{eq:Ampere3}
\end{equation}%
and the Higgs field equation reads%
\begin{eqnarray}
0 &=&\omega \left( D_{j}D_{j}\phi \right) +\left( \partial _{j}w\right)
D_{j}\phi  \label{eq:Higgs3} \\[0.15cm]
&&-\frac{\partial \omega }{\partial \phi ^{\ast }}\left\vert D_{j}\phi
\right\vert ^{2}-\frac{B^{2}}{\ 2\mathcal{R}}\frac{\partial G}{\partial \phi
^{\ast }}-\frac{\partial W}{\partial \phi ^{\ast }}.  \notag
\end{eqnarray}%
In the last two equations $\mathcal{R}$ reads
\begin{equation}
\mathcal{R}=\left( 1+\frac{G}{\beta ^{2}}B^{2}\right) ^{1/2}.  \label{RR}
\end{equation}

The energy of the system, in static regime and in the gauge $A_{0}=0$, is
given by
\begin{equation}
E=\int \!d^{2}x\left[ \beta ^{2}\left( \mathcal{R}-V\right) +\omega
\left\vert D_{k}\phi \right\vert ^{2}\right] ,  \label{t00}
\end{equation}%
and will be nonnegative whenever the condition $\mathcal{R}\geq V$ is
satisfied.

To proceed with the BPS formalism, we use the identities (\ref{iden}) and (%
\ref{eqw7}) such that the Eq. (\ref{t00}) becomes
\begin{eqnarray}
E &=&2\pi \!\int \!d^{2}x\Bigg[\pm ev^{2}B+\omega |D_{\pm }\phi |^{2}\pm
\frac{1}{2\lambda }\epsilon _{ik}\partial _{i}\left( \omega J_{k}\right)
\notag \\
&&\hspace{1.5cm}+\frac{\mathcal{R}}{2G}\left( \frac{G}{\mathcal{R}}B\mp
\sqrt{2U^{(\lambda )}}\right) ^{2}  \label{E2} \\
&&\hspace{1.5cm}+\beta ^{2}\left( \mathcal{R}-V\right) -\frac{1}{2}\frac{%
GB^{2}}{\mathcal{R}}-\frac{\mathcal{R}}{G}U^{(\lambda )}\Bigg].  \notag
\end{eqnarray}%
We have introduced the potential $U^{(\lambda )}(|\phi |)$ given by Eq. (\ref%
{mhlambda}) with the aim to obtain the term proportional to the magnetic
field $ev^{2}B$.

The Bogomol'nyi procedure would be complete if we require that the third row
in (\ref{E2}) to be null, so we obtain
\begin{equation}
V+\frac{\mathcal{R}U^{(\lambda )}}{\beta ^{2}G}=\frac{1}{2}\mathcal{R}+\frac{%
1}{2\mathcal{R}}.  \label{XX}
\end{equation}%
It provides a relation between the functions $V$, $G$ and $\mathcal{R}$. We
here clarify that the Eq. (\ref{XX}) it is not arbitrary because, as we will
observe later, in the BPS limit it becomes equivalent to the condition on
the diagonal components of the energy-momentum tensor $T_{\mu \nu }$: $%
T_{11}+T_{22}=0$, proposed by Schaposnik and Vega \cite{Scha} to obtain
self-dual configurations.

Then, the condition (\ref{XX}) allows to write the energy (\ref{t00}) in the
Bogomol'nyi form,
\begin{eqnarray}
E &=&\int \!\!d^{2}x\left\{ \omega |D_{\pm }\phi |^{2}+\frac{\mathcal{R}}{2G}%
\left( \frac{G}{\mathcal{R}}B\mp \sqrt{2U^{(\lambda )}}\right) ^{2}\right.
\notag \\
&&\hspace{1.0cm}\pm ev^{2}B\pm \frac{1}{2\lambda }\epsilon _{ik}\partial
_{i}\left( \omega J_{k}\right) \Bigg\} .  \label{E4}
\end{eqnarray}

Under suitable boundary conditions, the integration of the total derivative
in Eq. (\ref{E4}) gives null contribution to the energy. Hence, it becomes
clear that the energy possess a lower bound
\begin{equation}
E\geq ev^{2}\left\vert \Phi \right\vert ,  \label{eq:bound}
\end{equation}%
with $\Phi $ the total magnetic flux. Such a lower bound is saturated when
the fields satisfy the BPS or self-dual equations
\begin{equation}
D_{\pm }\phi =0,  \label{BPS2_0}
\end{equation}%
\begin{equation}
\frac{G}{\mathcal{R}}B=\pm ev^{2}\left( 1-\frac{\left\vert \phi \right\vert
^{2\lambda }}{v^{2\lambda }}\right) .  \label{BPS2_1}
\end{equation}

\begin{figure}[]
\centering\includegraphics[width=8.6cm]{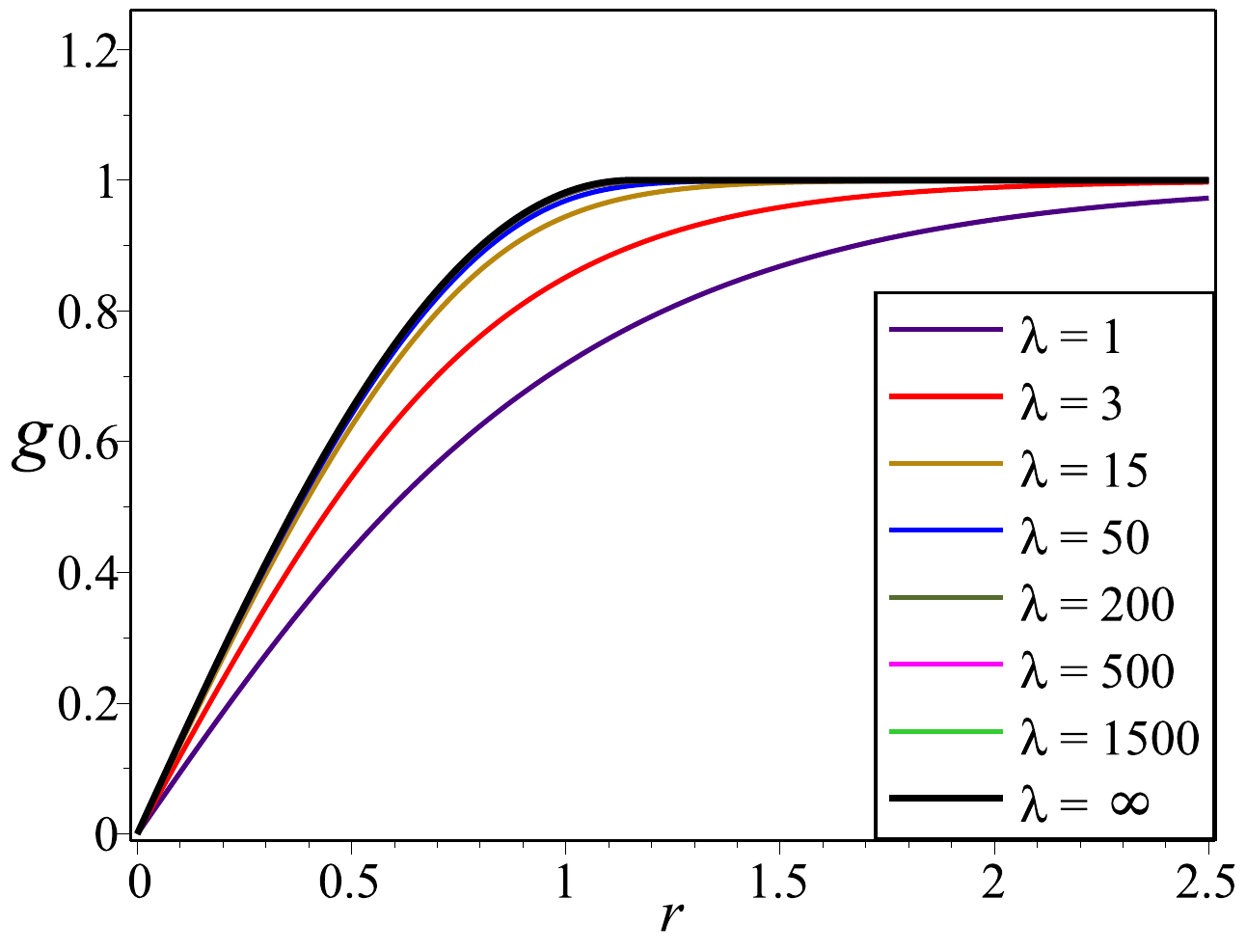}\vspace{0.18cm} %
\centering\includegraphics[width=8.6cm]{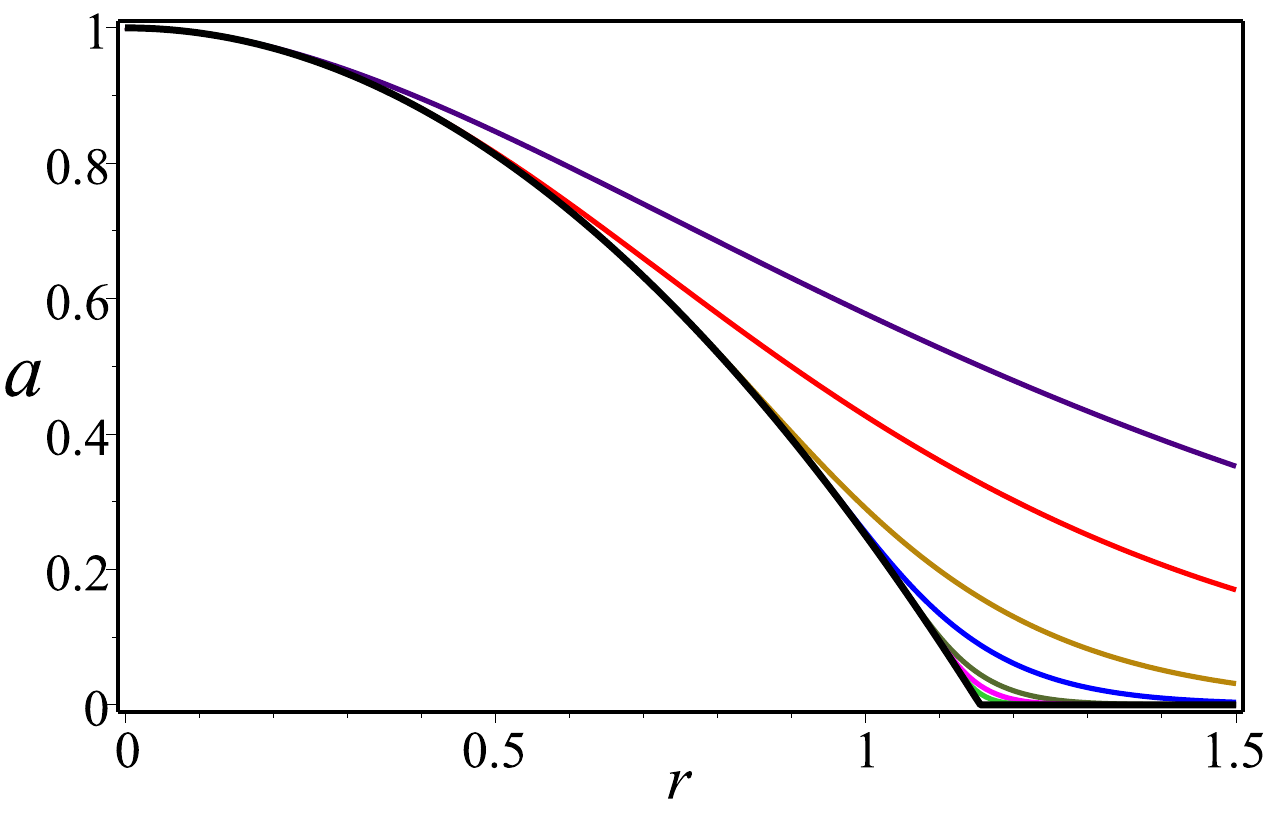}
\caption{The profiles $g(r)$ (upper) and $a(r)$ (lower) coming from the
generalized Born-Infeld-Higgs model (\protect\ref{eq:action}) with $G(g)=1$
and $\protect\omega (g)=\protect\lambda g^{2\protect\lambda -2}$. Observe that $\lambda=1$ (indigo lines) represents the usual BIH model and the true compacton solution is given by $\lambda=\infty$ (black lines).}
\label{fig03}
\end{figure}

\begin{figure}[]
\centering\includegraphics[width=8.6cm]{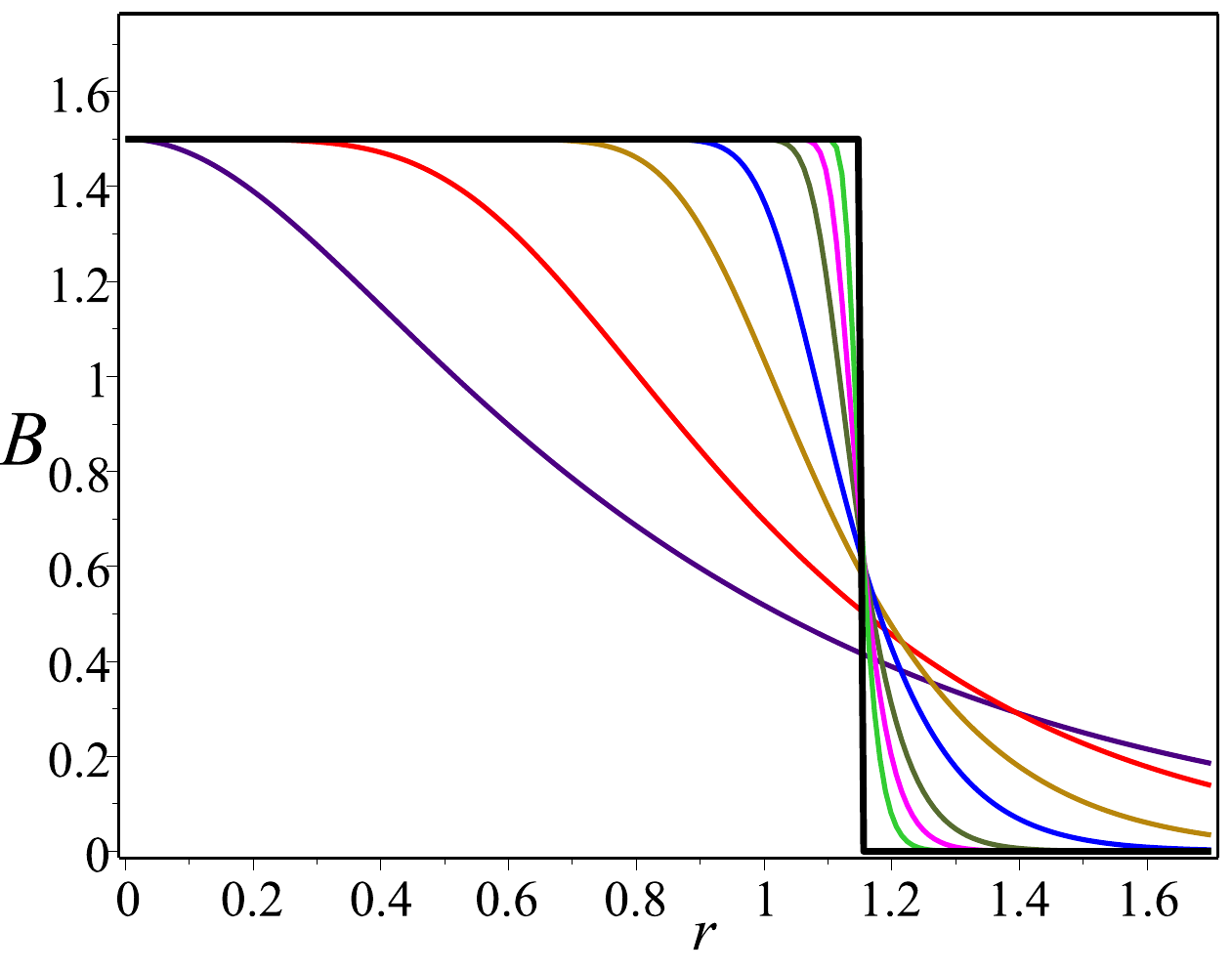}\vspace{0.23cm} %
\centering\includegraphics[width=8.6cm]{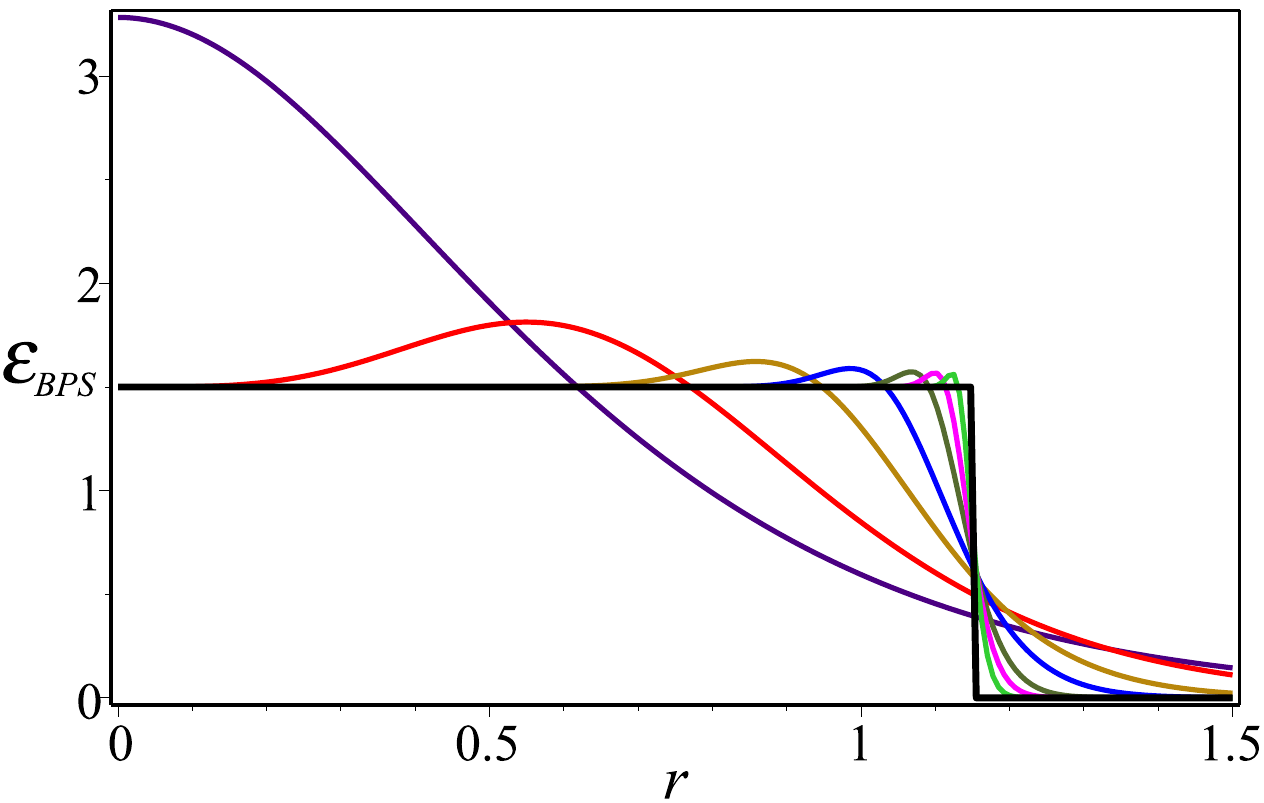}
\caption{The magnetic field $B(r)$ (upper) and the BPS energy density $\protect\varepsilon _{_{BPS}}(r)$ (lower) coming from the generalized
Born-Infeld-Higgs model (\protect\ref{eq:action}) with $G(g)=1$ and $\protect\omega (g)=\protect\lambda g^{2\protect\lambda -2}$. Observe that $\lambda=1$ (indigo lines) represents the usual BIH model and the true compacton solution is given by $\lambda=\infty$ (black lines).}
\label{fig04}
\end{figure}

By using the BPS equations in Eq. (\ref{XX}) we compute the self-dual
potential $V(|\phi |)$,
\begin{equation}
V=\frac{1}{\mathcal{R}}=\sqrt{1-\frac{2U^{(\lambda )}}{\beta ^{2}G}}
\end{equation}%
This way the second BPS equation (\ref{BPS2_1}) becomes%
\begin{equation}
B=\pm \frac{ev^{2}}{GV}\left( 1-\frac{\left\vert \phi \right\vert ^{2\lambda
}}{v^{2\lambda }}\right) .
\end{equation}%
By using the BPS equations in (\ref{t00}) we find the BPS energy density is
given by
\begin{equation}
\varepsilon _{_{BPS}}=\frac{2U^{(\lambda )}}{GV}+\lambda \frac{|\phi
|^{2\lambda -2}}{v^{2\lambda -2}}\left\vert D_{k}\phi \right\vert ^{2},
\end{equation}%
it will be positive-definite $\lambda >0$.

\subsection{Born-Infeld-Higgs effective compact vortices for $\protect\lambda$ finite}

In Ref. \cite{D2} it was explored the existence of effective compact vortex
solutions but the self-dual ones were not found. In this section we show the
existence of such self-dual effective compact solutions in BIH model.
Without loss of generality, we perform the study by considering $G(|\phi
|)=1 $ in the Lagrangian density (\ref{eq:action}).

The searching for vortex solutions is made by means of the vortex \textit{%
Ansatz} introduced in the Eq. (\ref{ax2}). Thus, the BPS equations (\ref%
{BPS2_0}) and (\ref{BPS2_1}) read
\begin{eqnarray}
g^{\prime } &=&\pm \frac{ag}{r},  \label{bih1s} \\[0.2cm]
-\frac{a^{\prime }}{r} &=&\pm \frac{e^{2}v^{2}\left( 1-g^{2\lambda }\right)
}{\sqrt{1-\displaystyle\frac{e^{2}v^{4}\left( 1-g^{2\lambda }\right) ^{2}}{%
\beta ^{2}}}},  \label{bih2s}
\end{eqnarray}

The behavior of the profiles $g(r)$ and $a(r)$ when $r\rightarrow 0$ is
determined by solving the self-dual equations (\ref{bih1s}) and (\ref{bih2s}%
), so we have
\begin{eqnarray}
g(r) &\approx &C_{n}r^{n}+...,  \label{Assym_g_BIH} \\[0.06in]
a(r) &\approx &n-\frac{e^{2}v^{2}\beta }{2\sqrt{\beta ^{2}-e^{2}v^{4}}}%
r^{2}+...,\quad \quad  \label{Assym_a_BIH}
\end{eqnarray}
Similarly, the behavior of the profiles for $r\rightarrow \infty $ is
\begin{eqnarray}
g(r) &\approx &1-\frac{C_{\infty }}{\sqrt{r}}e^{-mr}, \\[0.15cm]
a(r) &\approx &{m}C_{\infty }\sqrt{r}e^{-mr},
\end{eqnarray}%
where $m$, the self-dual mass, is given by
\begin{equation}
m=ev\sqrt{2\lambda }
\end{equation}%
it is exactly the same obtained for the generalized MH model analyzed in the
previous section.

The BPS energy density for the self-dual vortices reads
\begin{equation}
\varepsilon _{_{BPS}}=\displaystyle\frac{e^{2}v^{4}\left( 1-g^{2\lambda
}\right) ^{2}}{\sqrt{1-\displaystyle\frac{e^{2}v^{4}\left( 1-g^{2\lambda
}\right) ^{2}}{\beta ^{2}}}}+2\lambda v^{2}g^{2\lambda -2}\left( \frac{ag}{r}%
\right) ^{2},
\end{equation}%
it will be positive-definite and finite for $\lambda \geq 1$.

\subsection{Born-Infeld-Higgs compactons for $\protect\lambda=\infty$}

In the limit $\lambda \rightarrow \infty $, the BPS equations (\ref{bih1s})
and (\ref{bih2s}) read
\begin{eqnarray}
g^{\prime }&=&\pm \frac{ag}{r},  \label{bihcp1} \\[0.2cm]
-\frac{a^{\prime }}{r}&=&\pm \frac{e^{2}v^{2}\Theta (1-g)}{\displaystyle%
\sqrt{1-\frac{e^{2}v^{4}}{\beta ^{2}}}},  \label{bihcp2}
\end{eqnarray}%
with the profiles $g(r)$ and $a(r)$ satisfying the boundary conditions
(\ref{bcx1c}) and (\ref{bcx2c}).

By solving the BPS  compacton  equations for the BIH model, we obtain also
analytical solutions
\begin{eqnarray}
g^{(\infty )}(r)&=&\left(\frac{r}{r_{c}}\right)^n \exp \left[ \frac{n}{2}\left( 1-%
\frac{r^{2}}{r_{c}^{2}}\right) \right]\!  \Theta (r_{c}-r)\quad\\[0.2cm]
&&+\Theta (r-r_{c}), \nonumber\\[0.2cm]
a^{(\infty )}(r)&=&n\left( 1-\frac{r^{2}}{r_{c}^{2}}\right) \Theta \left(
r_{c}-r\right) ,
\end{eqnarray}%
where the radial distance $r_{c}$ now is given by
\begin{equation}
r_{c}=\frac{\sqrt{2n}}{\left\vert ev\right\vert }\left( 1-\frac{e^{2}v^{4}}{%
\beta ^{2}}\right) ^{1/4}.
\end{equation}

The magnetic field and BPS energy density profiles of the Born-Infeld-Higgs compacton are
\begin{eqnarray}
B^{(\infty )}(r) &=&ev^{2}\left( 1-\frac{e^{2}v^{4}}{\beta ^{2}}\right)
^{-1/2}\Theta (r_{c}-r), \\[0.2cm]
\varepsilon _{_{BPS}}^{(\infty )}(r) &=&e^{2}v^{4}\left( 1-\frac{e^{2}v^{4}}{%
\beta ^{2}}\right) ^{-1/2}\Theta (r_{c}-r).
\end{eqnarray}

In order to compute the numerical solutions we choose the upper signs in
equations (\ref{bih1s}) and (\ref{bih2s}), $e=1$, $v=1$, $\beta =3/\sqrt{5}$
and winding number $n=1$.  Similarly to the MH model, the effective compacton
behavior appears for sufficiently large values of $\lambda $, see Figs.
\ref{fig03} and \ref{fig04}. The true Born-Infeld-Higgs compacton arising for
$\lambda=\infty$ also are depicted (see black line profiles) in Figs.
\ref{fig03} and \ref{fig04}.

\section{The Chern-Simons-Higgs case}

In this section we apply the same formalism to construct self-dual solutions
in the generalized Abelian Chern-Simons-Higgs model. Physics in two
spatial dimensions is closely linked to CS theory, which contains
theoretical novelties besides practical application in various phenomena of
condensed matter, such as the physics of Anyons and it is related with the
fractional quantum Hall effect \cite{Ezawa}. It can be found an extensive
literature about CS theory, some of the pioneer papers concerning
topological and non-topological solutions as well as relativistic and
non-relativistic models can be found in \cite{hor,JW,JP,JLW}. Also exists a
close connection between CS theory and supersymmetry. This connection was
firstly demonstrated in \cite{LLW}, where from a $N=2$\ supersymmetric
extension of CS model it was found the specific potential for the
Bogomol'nyi equations, which arise naturally.

The generalized Chern-Simons-Higgs model is described by the following Lagrangian density
\begin{equation}
\mathcal{L}=\frac{\kappa }{4}\epsilon ^{\mu \nu \rho }A_{\mu }F_{\nu \rho
}+\omega (|\phi |)|D_{\mu }\phi |^{2}-V(|\phi |),  \label{Acg1}
\end{equation}%
with the function $\omega \left( \left\vert \phi \right\vert \right) $ given
by Eq. (\ref{omegaa}). The gauge field equation to be
\begin{equation}
\frac{\kappa }{2}\epsilon ^{\mu \alpha \beta }F_{\alpha \beta }-e\omega
J^{\mu }=0~,  \label{ge0}
\end{equation}%
and the Gauss law reads%
\begin{equation*}
\kappa B=e\omega J_{0}.
\end{equation*}
It is clear that the electric charge density is $e\omega J_{0}$ whose
integration performed via the Gauss law gives
\begin{equation}
Q=\int \!d^{2}x\,e\omega J_{0}=\kappa \int \!d^{2}x~B=\kappa \Phi .
\end{equation}%
So such as it\ happens in usual CSH model, the electric charge is nonnull
and proportional to the magnetic so\ the field configurations always will be
electrically charged.

These are the stationary points of the energy which for the static field
configuration reads
\begin{equation}
E=\!\!\int \!d^{2}x\!\left[ -\kappa A_{0}B-e^{2}\omega A_{0}^{2}|\phi
|^{2}+\omega |D_{i}\phi |^{2}+V(|\phi |)\right] .  \label{EJP}
\end{equation}%
\textbf{From the static Gauss law,} we obtain the relation
\begin{equation}
A_{0}=-\frac{\kappa }{2e^{2}}\frac{B}{\omega |\phi |^{2}},  \label{A0}
\end{equation}%
which substituted in Eq. (\ref{EJP}) leads to the following expression for
the energy:
\begin{equation}
E=\int \!d^{2}x\left[ \frac{\kappa ^{2}}{4e^{2}}\frac{B^{2}}{\omega |\phi
|^{2}}+\omega |D_{i}\phi |^{2}+V(|\phi |)\right] .  \label{EJP1}
\end{equation}

We now use the identities (\ref{iden}) and (\ref{eqw7}) in Eq. (\ref{EJP})
such that the energy becomes
\begin{eqnarray}
E &=&\!\int \!\!d^{2}x\left[ \frac{\kappa ^{2}}{4e^{2}}\frac{B^{2}}{\omega
|\phi |^{2}}+V(|\phi |)+\omega |D_{\pm }\phi |^{2}\right.  \label{EJP3} \\
&&\hspace{1.25cm}\left. \pm ev^{2}\frac{|\phi |^{2\lambda }}{v^{2\lambda }}%
B\pm \frac{1}{2\lambda }\epsilon _{ik}\partial _{i}\left( \omega
J_{k}\right) \right] .  \notag
\end{eqnarray}

After some manipulation the energy can be expressed almost in the
Bogomol'nyi form
\begin{eqnarray}
E &=&\!\int \!d^{2}x\left[ \frac{\kappa ^{2}}{4e^{2}}\frac{1}{|\phi
|^{2}\omega }\left( B\mp \frac{2e\left\vert \phi \right\vert }{\kappa }\sqrt{%
\omega V}\right) ^{2}\right.  \notag \\
&&\hspace{1.25cm}+\omega |D_{\pm }\phi |^{2}\pm \frac{1}{2\lambda }\epsilon
_{ik}\partial _{i}\left( \omega J_{k}\right)  \notag \\
&&\hspace{1.25cm}\left. \pm B\left( \frac{\kappa }{e|\phi |}\sqrt{\frac{V}{%
\omega }}+ev^{2}\frac{|\phi |^{2\lambda }}{v^{2\lambda }}\right) \right] ~.
\label{EPJ4}
\end{eqnarray}%
We observe that the Bogomol'nyi procedure will be complete if the term
multiplying the magnetic field is a constant, i.e.,%
\begin{equation}
\frac{\kappa }{e|\phi |}\sqrt{\frac{V}{\omega }}+ev^{2}\frac{|\phi
|^{2\lambda }}{v^{2\lambda }}=ev^{2},
\end{equation}%
such a condition allows to determine the self-dual potential $V(|\phi |)$ to
be
\begin{equation}
V(|\phi |)=\lambda \frac{e^{4}v^{6}}{\kappa ^{2}}\frac{|\phi |^{2\lambda }}{%
v^{2\lambda }}\left( 1-\frac{\left\vert \phi \right\vert ^{2\lambda }}{%
v^{2\lambda }}\right) ^{2}.  \label{potsbps0}
\end{equation}%
We can see that for $\lambda =1$, the $|\phi |^{6}$-potential of the usual
Chern-Simon-Higgs model is recovered.

Hence, the energy (\ref{EPJ4}) reads%
\begin{eqnarray}
E &=&\!\!\int \! d^{2}x\left\{ \pm ev^{2}B\pm \frac{1}{2\lambda }\epsilon
_{ik}\partial _{i}\left( \omega J_{k}\right) +\omega |D_{\pm }\phi
|^{2}\right.\quad \\
&&\hspace{0.7cm}\left. +\frac{\kappa ^{2}}{4e^{2}|\phi |^{2}\omega }\left[
B\mp \frac{2e^{2}v^{2}}{\kappa ^{2}}\lambda \frac{|\phi |^{2\lambda }}{%
v^{2\lambda }}\sqrt{2U^{(\lambda )}}\right] ^{2}\right\} .  \notag
\end{eqnarray}%
We see that under appropriated boundary conditions the total derivative
gives null contribution to the energy. Then, the energy is bounded below by
a multiple of the magnetic flux magnitude
\begin{equation}
E\geq \pm ev^{2}\!\!\int \!\!d^{2}xB=ev^{2}|\Phi |.
\end{equation}%
This bound is saturated by fields satisfying the first-order Bogomol'nyi or
self-dual equations \cite{BPS}
\begin{equation}
D_{\pm }\phi =0,  \label{BPS1}
\end{equation}%
\begin{equation}
B=\pm \frac{2e^{3}v^{4}}{\kappa ^{2}}\lambda \frac{|\phi |^{2\lambda }}{%
v^{2\lambda }}\left( 1-\frac{\left\vert \phi \right\vert ^{2\lambda }}{%
v^{2\lambda }}\right) .  \label{BPS2}
\end{equation}%
In order the magnetic field be nonsingular at origin, it is required that
the $\lambda >0$.

By using the BPS equation in (\ref{EJP1}) we find the energy density is
given by
\begin{equation}
\varepsilon _{_{BPS}}=2\lambda \frac{e^{4}v^{6}}{\kappa ^{2}}\frac{|\phi
|^{2\lambda }}{v^{2\lambda }}\left( 1-\frac{\left\vert \phi \right\vert
^{2\lambda }}{v^{2\lambda }}\right) ^{2}+\lambda \frac{|\phi |^{2\lambda -2}%
}{v^{2\lambda -2}}\left\vert D_{k}\phi \right\vert ^{2},
\end{equation}%
it will be positive-definite $\lambda >0$.

\begin{figure}[]
\centering\includegraphics[width=8.6cm]{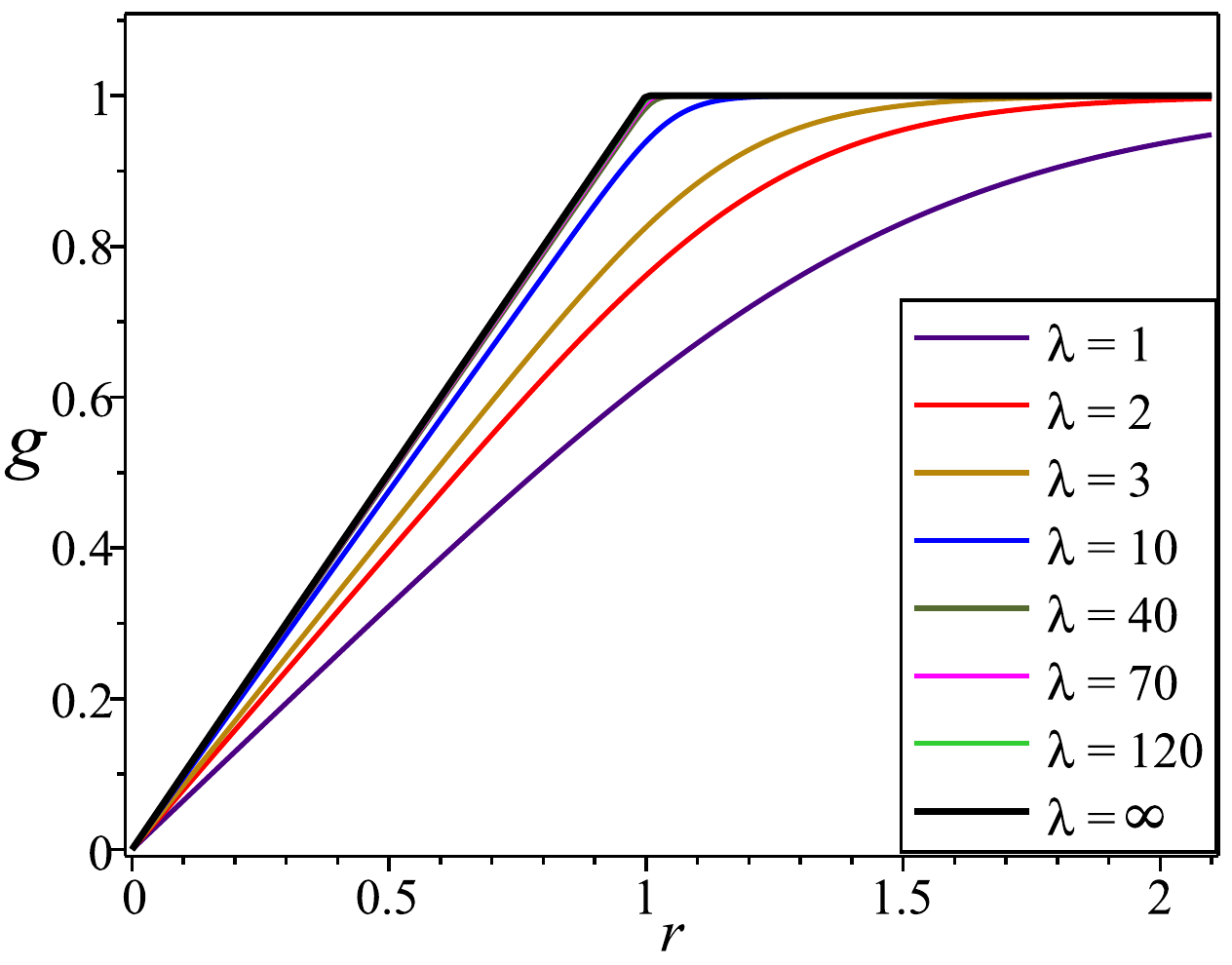} \centering%
\includegraphics[width=8.6cm]{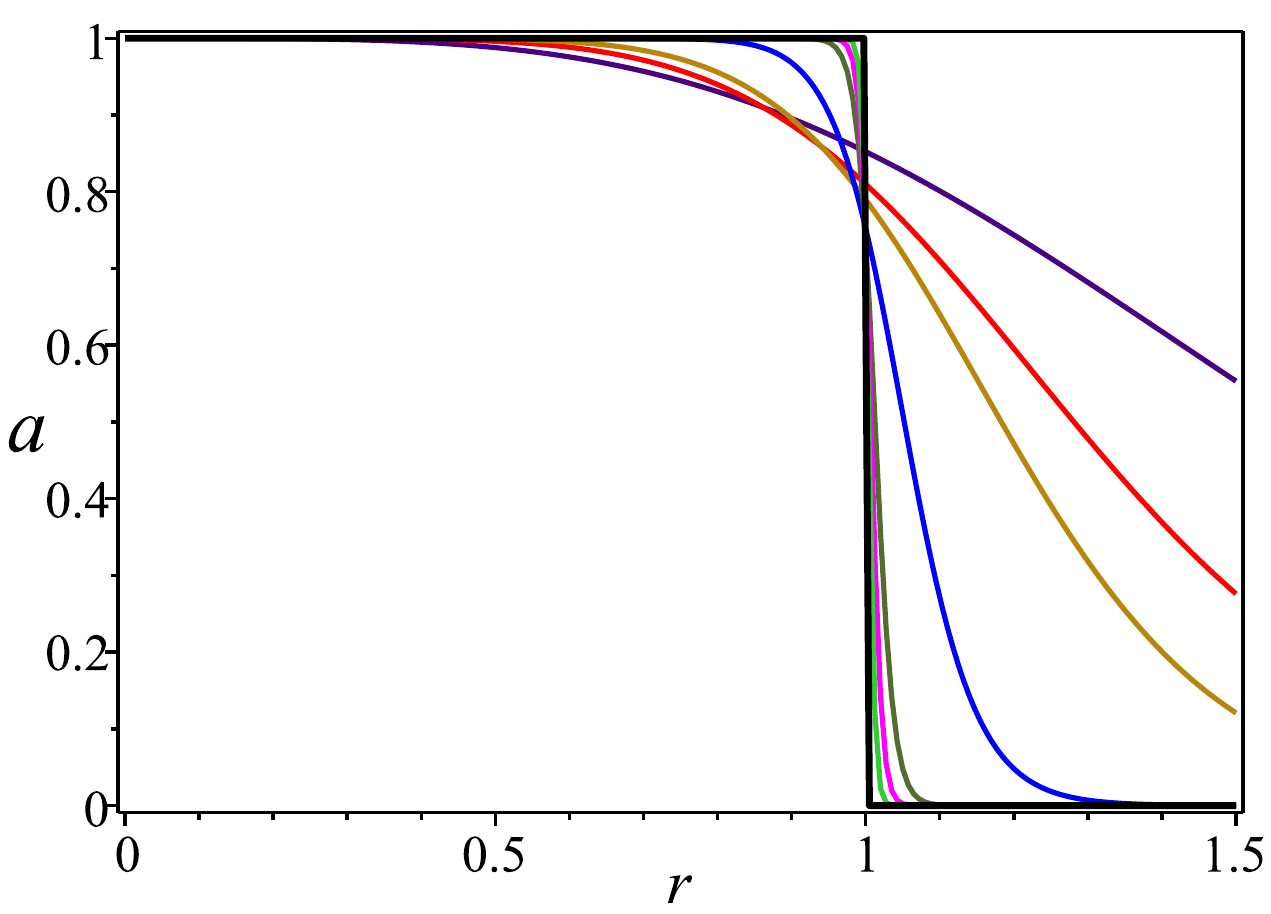}
\caption{The profiles $g(r) $ (upper) and $a(r) $ (lower) coming from the
generalized Chern-Simons-Higgs model (\protect\ref{Acg1}) with $\protect\omega (g)=\protect\lambda g^{2\protect\lambda -2}$. Observe that $\lambda=1$ (indigo lines) represents the usual CSH model and the true compacton solution is given by $\lambda=\infty$ (vertical black lines representing the $\delta$-Dirac function).}
\label{fig05}
\end{figure}

\begin{figure}[]
\centering\includegraphics[width=8.6cm]{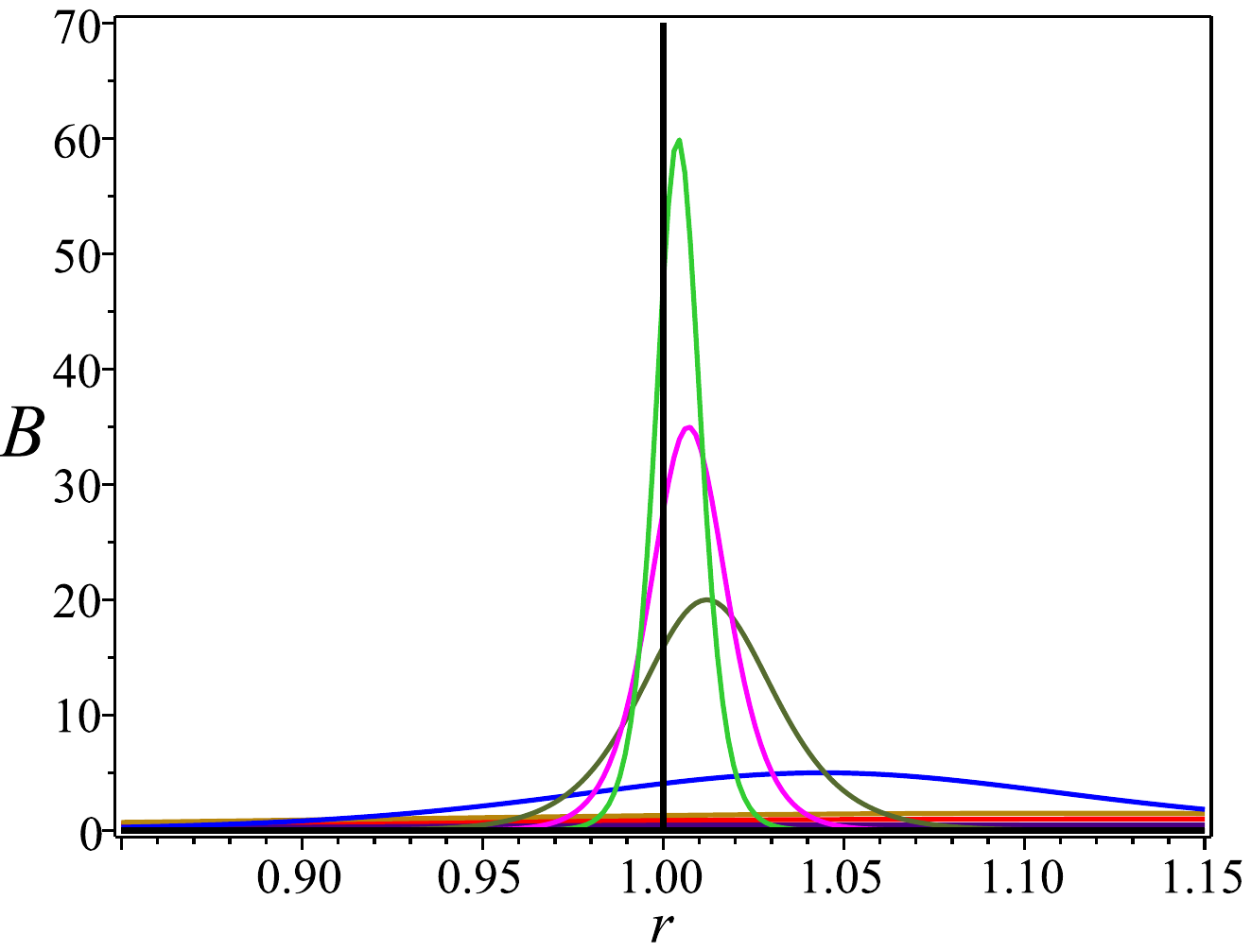} \centering%
\includegraphics[width=8.6cm]{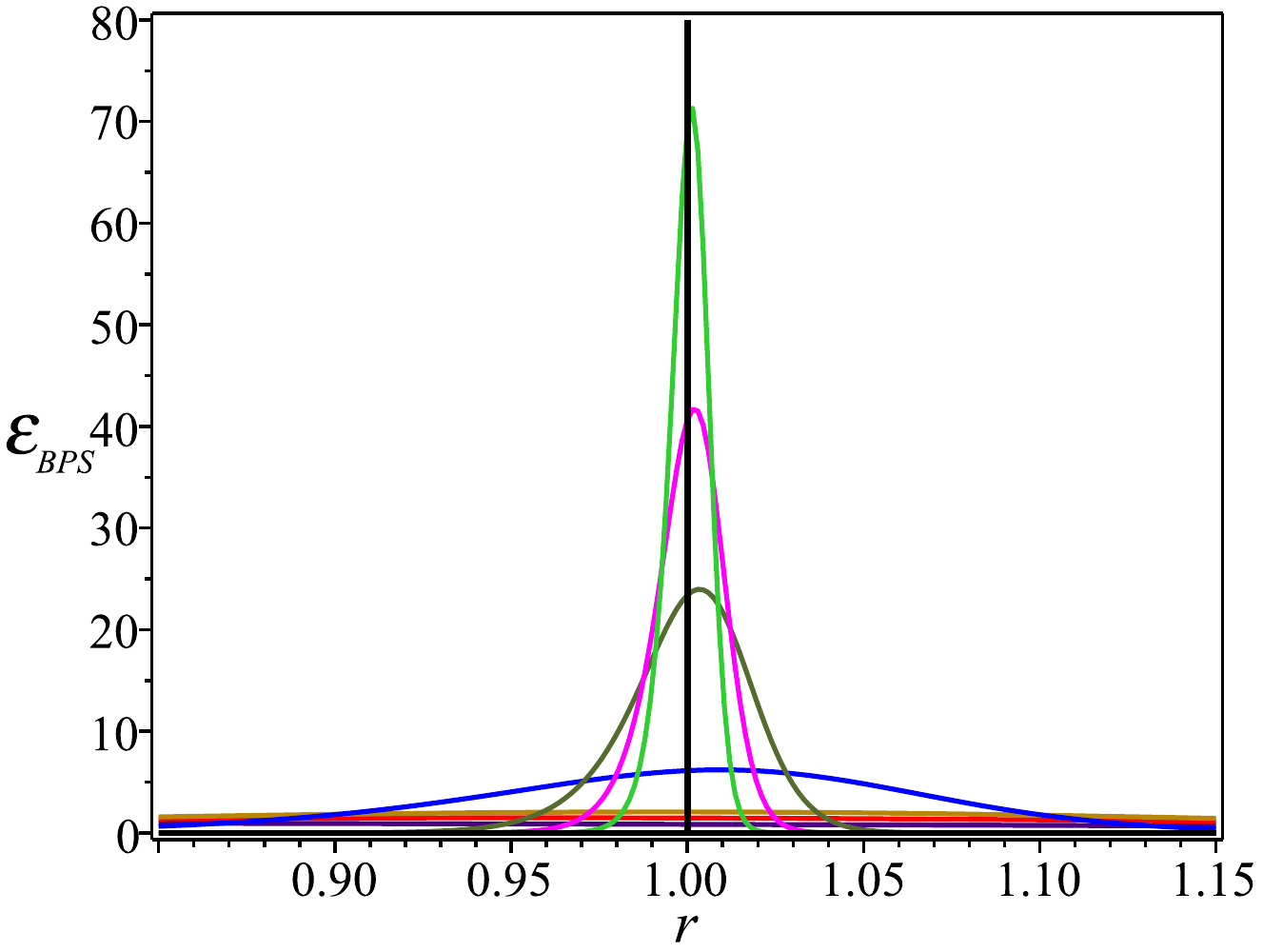}
\caption{The magnetic field $B(r)$ (upper) and the BPS energy density $\protect\varepsilon _{_{BPS}}(r)$ (lower) coming from the generalized
Chern-Simons-Higgs model (\protect\ref{Acg1}) with $\protect\omega (g)=\protect\lambda g^{2\protect\lambda -2}$. Observe that $\lambda=1$ (indigo lines) represents the usual CSH model and the true compacton solution is given by $\lambda=\infty$ (vertical black lines representing the $\delta$-Dirac function).} \label{fig06}
\end{figure}

\subsection{Chern-Simons-Higgs effective compact vortices for $\protect\lambda$ finite}

The BPS equations (\ref{BPS1}) and (\ref{BPS2}) read
\begin{eqnarray}
g^{\prime }&=&\pm \frac{ag}{r},  \label{bcs1} \\[0.2cm]
-\frac{a^{\prime }}{r}&=&\pm \frac{2e^{4}v^{4}}{\kappa ^{2}}\lambda
g^{2\lambda }(1-g^{2\lambda }),  \label{bcs2}
\end{eqnarray}

The behavior of $g(r)$ and $a(r)$ near the boundaries can be easily
determined by solving the self-dual equations (\ref{bcs1}) and (\ref{bcs2})
around the boundary values (\ref{bcx1}) and (\ref{bcx2}). Thus, for $%
r\rightarrow 0$, the profile functions behave as
\begin{eqnarray}
g(r) &\approx &C_{n}r^{n}+...,  \label{Assym_g_CSH} \\[0.2cm]
a(r) &\approx &n\mp \frac{\lambda e^{4}v^{4}}{\left( n\lambda +1\right)
\kappa ^{2}}\left( C_{n}\right) ^{2\lambda }r^{2n\lambda +2}+...,
\label{Assym_a_CSH}
\end{eqnarray}%
where the constant $C_{n}>0$ is determined only numerically.

On the other hand, when $r\rightarrow \infty $ they behave as
\begin{eqnarray}
g(r) &\approx &1-\frac{C_{\infty }}{\sqrt{r}}e^{-mr}, \\[0.15cm]
a(r) &\approx &C_{\infty }{m}\sqrt{r}e^{-mr},
\end{eqnarray}%
with the constant $C_{\infty }$ computed numerically and $m$ being the
self-dual mass
\begin{equation}
m=\frac{2\lambda e^{2}v^{2}}{\kappa }.
\end{equation}%
Observe for $\lambda=1$ it becomes the one  of the usual self-dual
Chern-Simons-Higgs bosons.

The BPS energy density for the self-dual vortices is given by
\begin{equation}
\varepsilon _{_{BPS}}=2\lambda \frac{e^{4}v^{6}}{\kappa ^{2}}g^{2\lambda
}\left( 1-g^{2\lambda }\right) ^{2}+2\lambda v^{2}g^{2\lambda -2}\left(
\frac{ag}{r}\right) ^{2},
\end{equation}%
and it will be positive-definite and finite for $\lambda \geq 1$.

\subsection{Chern-Simons-Higgs compactons for $\protect\lambda=\infty$}

From Eqs. (\ref{bcs1}) and (\ref{bcs2}) we obtain the BPS equations for the
Chern-Simons-Higgs compactons
\begin{eqnarray}
g^{\prime }&=&\pm \frac{ag}{r}, \\[0.2cm]
-\frac{a'}{r}&=&\frac{2 e^4 v^4}{\kappa^2}\delta(1-g),
\end{eqnarray}
with the boundary condition given in Eqs. (\ref{bcx1c}) and (\ref{bcx2c}).

By solving the BPS compacton equations for $n>0$, we obtain the analytic profiles
\begin{eqnarray}
g^{(\infty )}(r) &=&\left(\frac{r}{r_{c}}\right)^n \Theta (r_{c}-r)+\Theta (r-r_{c}),
\\[0.2cm]
a^{(\infty )}(r) &=&n\Theta \left( r_{c}-r\right) .
\end{eqnarray}%
The radial distance $r_{c}$ is calculated to be
\begin{equation}
r_c=\frac{n\left\vert \kappa \right\vert }{e^{2}v^{2}}.
\end{equation}
The magnetic field and BPS energy density of the Chern-Simons-Higgs  compacton
are
\begin{eqnarray}
B^{(\infty )}(r) &=&\frac{n}{er_{c}}\delta (r_{c}-r), \\[0.2cm]
\varepsilon _{_{BPS}}^{(\infty )}(r) &=&\frac{nv^{2}}{r_{c}}\delta (r_{c}-r).
\end{eqnarray}

In order to compute the numerical solutions we choose the upper signs in
equations (\ref{bcs1}) and (\ref{bcs2}), $e=1$, $v=1$, $\kappa =1$ and winding
number $n=1$. From the numerical analysis, we can see the appearing of the
effective compacton behavior for not very large values of $\lambda $ such it is
explicitly shown in Figs. \ref{fig05} and \ref{fig06}. An interesting feature
of the Chern-Simons-Higgs effective compact vortices is the enhancement of the
ring shape (inclusive for $n=1$), for increasing values of $\lambda$ in the
profiles for the magnetic field and the BPS energy density (see Fig.
\ref{fig06}).  The analytic CSH compacton structures appearing in
$\lambda=\infty$ are represented by the black line profiles in Fig. \ref{fig05}
and \ref{fig06}.

\section{The Maxwell-Chern-Simons-Higgs case}

Electrically charged vortices were first found in the Abelian-Higgs model by
S. K. Paul and A. Khare \cite{Khare}, where the Chern-Simons term was
included in the usual Maxwell-Higgs action. This was an ingenious manner to
avoid the temporal gauge, $A_{0}\neq 0$, coupling the electric charge
density to the magnetic field. This model has also been generalized by
multiplying a dielectric (scalar) function in the Maxwell kinetic term \cite%
{Ghosh,Bazeia2} yielding topological and not topological solutions
satisfying a Bogomol'nyi bound.

The generalized Maxwell-Chern-Simons-Higgs model is described by the
following Lagrangian density%
\begin{eqnarray}
\mathcal{L} &=&-\frac{G(|\phi |)}{4}F_{\mu \nu }F^{\mu \nu }+\frac{\kappa }{4%
}\epsilon ^{\mu \nu \rho }A_{\mu }F_{\nu \rho }  \label{Acg2} \\
&&+\omega (|\phi |)|D_{\mu }\phi |^{2}+\frac{G(|\phi |)}{2}\partial _{\mu
}N\partial ^{\mu }N  \notag \\
&&-e^{2}\omega (|\phi |)N^{2}|\phi |^{2}-V(|\phi |)~.  \notag
\end{eqnarray}%
where the function $\omega \left( \left\vert \phi \right\vert \right) $
given by Eq. (\ref{omegaa}). The gauge field equation reads
\begin{equation}
\partial _{\nu }\left( GF^{\nu \mu }\right) +\frac{\kappa }{2}\epsilon ^{\mu
\alpha \beta }F_{\alpha \beta }-e\omega J^{\mu }=0,
\end{equation}%
the static Gauss law is
\begin{equation}
\partial _{k}\left( G\partial _{k}A_{0}\right) -\kappa B=2e^{2}\omega
A_{0}\left\vert \phi \right\vert ^{2}.
\end{equation}

Similarly, the equation of motion of the Higgs field is
\begin{eqnarray}
0 &=&D_{\mu }\left( \omega D^{\mu }\phi \right) +\left( \frac{1}{4}F_{\mu
\nu }F^{\mu \nu }-\frac{1}{2}\partial _{\mu }N\partial ^{\mu }N\right) \frac{%
\partial G}{\partial \phi ^{\ast }}  \notag \\
&&-\frac{\partial \omega }{\partial \phi ^{\ast }}|D_{\mu }\phi
|^{2}+2e^{2}\omega N|\phi |^{2}+\frac{\partial V}{\partial \phi ^{\ast }}~.
\end{eqnarray}

Likewise than previous models, we are interested in time-independent soliton
solutions that ensure the finiteness of the action (\ref{Acg1}). These are
the stationary points of the energy which for the static field configuration
reads
\begin{eqnarray}
E &=&\!\!\int \!d^{2}x\!\left[ \frac{G}{2}B^{2}+\frac{G}{2}\left( \partial
_{j}A_{0}\right) ^{2}+e^{2}\omega \left( A_{0}\right) ^{2}|\phi |^{2}\right.
\\
&&\left. +\omega |D_{j}\phi |^{2}+\frac{G}{2}\left( \partial _{j}N\right)
^{2}+e^{2}\omega N^{2}|\phi |^{2}+V(|\phi |)\right] .  \notag
\end{eqnarray}%
To proceed, we use the identities (\ref{iden}) and (\ref{eqw7}) such that
the energy becomes
\begin{eqnarray}
E &=&\!\!\int \!d^{2}x\!\left[ \frac{G}{2}B^{2}+V(|\phi |)+\frac{G}{2}\left(
\partial _{j}A_{0}\right) ^{2}+\frac{G}{2}\left( \partial _{j}N\right)
^{2}\right.  \notag \\
&&+e^{2}\omega \left( A_{0}\right) ^{2}|\phi |^{2}+e^{2}\omega N^{2}|\phi
|^{2}  \label{mcsh2} \\
&&\left. +\omega |D_{\pm }\phi |^{2}\pm ev^{2}\frac{|\phi |^{2\lambda }}{%
v^{2\lambda }}B\pm \frac{1}{2\lambda }\epsilon _{ik}\partial _{i}\left(
\omega J_{k}\right) \ \right] .  \notag
\end{eqnarray}%
After some algebraic manipulations, it can be expressed by
\begin{eqnarray}
E &=&\!\!\int \!d^{2}x\!\left[ \omega |D_{\pm }\phi |^{2}+\frac{G}{2}\left(
B\mp \sqrt{\frac{2V}{G}}\right) ^{2}\right.  \notag \\
&&+\frac{G}{2}\left( \partial _{j}A_{0}\pm \partial _{j}N\right)
^{2}+e^{2}\omega |\phi |^{2}\left( A_{0}\pm N\right) ^{2}  \notag \\
&&\pm B\left( \sqrt{2GV}+ev^{2}\frac{|\phi |^{2\lambda }}{v^{2\lambda }}%
+\kappa N\right)  \notag \\
&&\left. \mp \partial _{j}\left( NG\partial _{j}A_{0}\right) \pm \frac{1}{%
2\lambda }\epsilon _{ik}\partial _{i}\left( \omega J_{k}\right) \right] \ .
\label{mcsh3}
\end{eqnarray}
At this point, with the purpose the total energy to have a lower bound
proportional to the magnetic field, we chose the potential $V(|\phi |)$ to
be
\begin{equation}
V(|\phi |)=\frac{1}{2G}\left( ev^{2}-ev^{2}\frac{\left\vert \phi \right\vert
^{2\lambda }}{v^{2\lambda }}-\kappa N\right) ^{2},  \label{mcsh4}
\end{equation}%
where $C=\lambda v^{2-2\lambda }$ in order to the vacuum expectation value
of the Higgs field be $|\phi |=v$. Hence, the energy (\ref{mcsh3}) reads%
\begin{eqnarray}
E &=&\int\!d^{2}x\left\{\pm ev^{2}B+\omega|D_{\pm}\phi|^{2}\frac{}{}\right.
\notag \\
&&+\frac{G}{2}\left[ B\mp \frac{1}{G}\left(ev^{2}-ev^{2}\frac{\left\vert
\phi \right\vert ^{2\lambda }}{v^{2\lambda }}-\kappa N\right) \right] ^{2}
\notag \\
&& +\frac{G}{2}\left( \partial _{j}A_{0}\pm \partial _{j}N\right)
^{2}+e^{2}\omega |\phi |^{2}\left( A_{0}\pm N\right) ^{2}  \notag \\
&&\left. \mp \partial _{j}\left( NG\partial _{j}A_{0}\right) \pm \frac{1}{%
2\lambda }\epsilon _{ik}\partial _{i}\left( \omega J_{k}\right)\right\}.
\label{mcsh5}
\end{eqnarray}
Under appropriated boundary conditions on the fields, the integration of the
total derivatives becomes null, then the total energy is bounded below by a
multiple of the magnetic flux magnitude
\begin{equation}
E\geq \pm ev^{2}\!\!\int \!\!d^{2}xB=ev^{2}|\Phi|. \label{enn}
\end{equation}

The lower-bound (\ref{enn}) is saturated by fields satisfying the first-order Bogomol'nyi or self-dual equations \cite{BPS}
\begin{equation}
D_{\pm }\phi =0,  \label{bps_mcsh1}
\end{equation}%
\begin{equation}
B=\pm \frac{ev^{2}}{G}\left( 1-\frac{\left\vert \phi \right\vert ^{2\lambda }%
}{v^{2\lambda }}\right) \mp \kappa \frac{N}{G}.  \label{bps_mcsh2}
\end{equation}%
\begin{equation}
\partial _{j}A_{0}\pm \partial _{j}N=0  \label{bps_mcsh3}
\end{equation}%
\begin{equation}
A_{0}\pm N=0  \label{bps_mcsh4}
\end{equation}

The condition $N=\mp A_{0}$ saturates the two last equations (\ref{bps_mcsh3}%
) and (\ref{bps_mcsh4}) so the self-dual solutions are obtained by solving
the following self-dual equations
\begin{equation}
D_{\pm }\phi =0,  \label{bps_mcsh1a}
\end{equation}%
\begin{equation}
B=\pm \frac{ev^{2}}{G}\left( 1-\frac{\left\vert \phi \right\vert ^{2\lambda }%
}{v^{2\lambda }}\right) +\kappa \frac{A_{0}}{G},  \label{bps_mcsh2a}
\end{equation}%
and the Gauss law%
\begin{equation}
\partial _{k}\left( G\partial _{k}A_{0}\right) -\kappa B=2e^{2}v^{2}\lambda
\frac{|\phi |^{2\lambda }}{v^{2\lambda }}A_{0}.  \label{gauss_3a}
\end{equation}

The BPS energy density is%
\begin{eqnarray}
\varepsilon _{_{BPS}} &=&GB^{2}+G\left( \partial _{j}A_{0}\right)
^{2}+\lambda \frac{|\phi |^{2\lambda -2}}{v^{2\lambda -2}}|D_{j}\phi |^{2}
\notag \\
&&+2e^{2}v^{2}\lambda \frac{|\phi |^{2\lambda }}{v^{2\lambda }}\left(
A_{0}\right) ^{2}.
\end{eqnarray}

\subsection{Maxwell-Chern-Simons-Higgs effective compact vortices for $\protect\lambda$ finite}

By considering $G(|\phi |)=1$ and using the \textit{Ansatz} (\ref{ax2}), the
BPS equations (\ref{bps_mcsh1a}) and (\ref{bps_mcsh2a}) read
\begin{eqnarray}
g^{\prime }&=&\pm \frac{ag}{r},  \label{mcshb1}\\[0.2cm]
-\frac{a^{\prime }}{r}&=&\pm e^{2}v^{2}(1-g^{2\lambda })+e\kappa A_{0},
\label{mcshb2}
\end{eqnarray}
and the Gauss law (\ref{gauss_3a}) becomes
\begin{equation}
\frac{1}{r}\left( rA_{0}^{\prime }\right) ^{\prime }-\kappa B=2\lambda
e^{2}v^{2}g^{2\lambda }A_{0}.  \label{mcshb3}
\end{equation}

We analyze the behavior of the profiles $g(r)$ and $a(r)$ and $A_{0}(r)$ at
boundaries. This way, for $r\rightarrow 0$, the profiles behave as
\begin{eqnarray}
g(r) &\approx &C_{n}r^{n}+...,  \label{MCSH_g_0} \\[0.08in]
a(r) &\approx &n-\frac{e[ev^{2}+\kappa A_{0}(0)]}{2}r^{2}\!+...,\quad \quad
\label{MCSH_a_0} \\[0.08in]
A_{0}(r) &\approx &A_{0}(0)+\frac{\kappa \lbrack ev^{2}+\kappa A_{0}(0)]}{4}%
r^{2}+...,  \label{MCSH_A0_0_1}
\end{eqnarray}%
with the constants $C_{n}>0$ and $A_{0}(0)$ are determined numerically for
every $n$.

The behavior at origin for $A_{0}(r)$ provides the boundary condition
\begin{equation}
A_{0}^{\prime }(0)=0.  \label{A0_origin}
\end{equation}

On the other hand, for large values of $r$ ($r\rightarrow \infty )$ they
have the Abrikosov-Nielsen-Olesen behavior,
\begin{eqnarray}
g(r) &\approx &1-\frac{C_{\infty }}{\sqrt{r}}e^{-mr}, \\[0.15cm]
a(r) &\approx &C_{\infty }{m}\sqrt{r}e^{-mr}, \\[0.15cm]
A_{0}(r) &\approx &-\frac{|\kappa |}{\kappa }\frac{m}{e}\frac{C_{\infty }}{%
\sqrt{r}}e^{-mr},  \label{A0inf}
\end{eqnarray}%
the constant $C_{\infty }$ is computed numerically and $m$ being the
self-dual mass,
\begin{equation}
m=\frac{1}{2}\sqrt{\kappa ^{2}+8\lambda e^{2}v^{2}}-\frac{|\kappa |}{2},
\end{equation}%
for $\lambda =1$, we recover self-dual mass of the usual
Maxwell-Chern-Simons-Higgs bosons.

In this way we obtain from (\ref{A0inf}) the boundary condition for $%
A_{0}(r) $ when $r\rightarrow \infty $:
\begin{equation}
A_{0}(\infty )=0.  \label{A0_infty}
\end{equation}

The BPS energy density of the self-dual vortices reads
\begin{eqnarray}
\varepsilon _{_{BPS}} &=&B^{2}+\left(A^{\prime }_{0}\right)
^{2}+2v^{2}\lambda g^{2\lambda -2}\left( \frac{ag}{r}\right) ^{2}  \notag \\%
[0.2cm]
&&+2e^{2}v^{2}\lambda g^{2\lambda }\left( A_{0}\right) ^{2},
\end{eqnarray}
being positive-definite and finite for $\lambda \geq 1$.

\subsection{Maxwell-Chern-Simons-Higgs compactons for $\protect\lambda=\infty$}

From Eqs. (\ref{mcshb1}), (\ref{mcshb2}), the limit $\lambda \rightarrow
\infty $\ provides the BPS\ equation for the compacton configurations
\begin{eqnarray}
g^{\prime }&=&\pm \frac{ag}{r},\\[0.2cm]
-\frac{a^{\prime }}{r}&=&\pm e^{2}v^{2}\Theta (1-g)+e\kappa A_{0}.
\end{eqnarray}%
The compacton Gauss law obtained from Eq. (\ref{mcshb3}) becomes%
\begin{equation}
A_{0}^{\prime}+\frac{\kappa }{e}\frac{\left( a-n\right) }{r}=0.
\end{equation}

\begin{figure}[]
\centering\includegraphics[width=8.6cm]{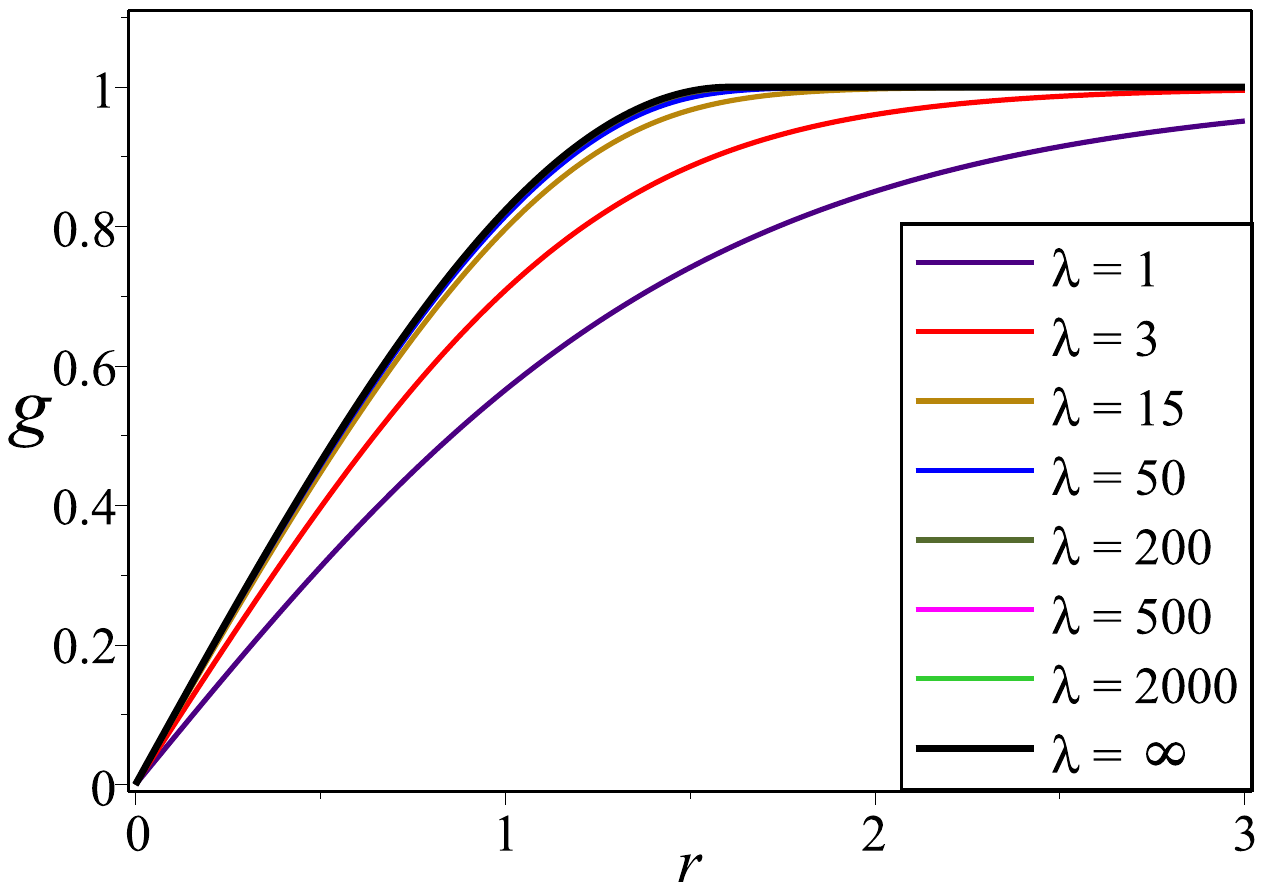} \centering%
\includegraphics[width=8.6cm]{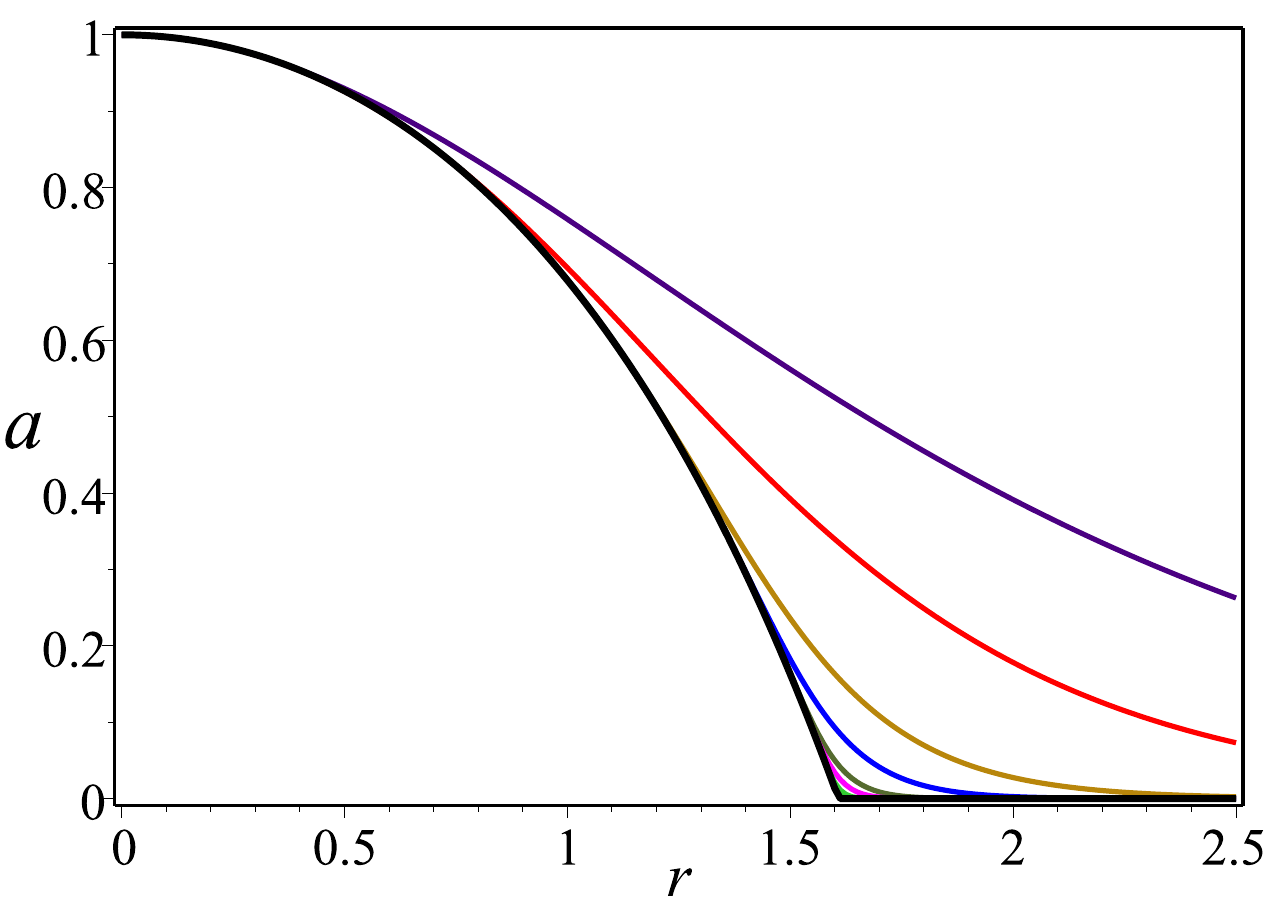}
\caption{The profiles $g(r) $ (upper) and $a(r) $ (lower) coming from
generalized Maxwell-Chern-Simons-Higgs model (\protect\ref{Acg2}) with $G(g)=1$ and $\protect\omega(g)=\protect\lambda g^{2\protect\lambda-2}$. Observe that $\lambda=1$ (indigo lines) represents the usual MCSH model and the true compacton solution is given by $\lambda=\infty$ (black lines).}
\label{fig07}
\end{figure}

\begin{figure}[]
\centering\includegraphics[width=8.6cm]{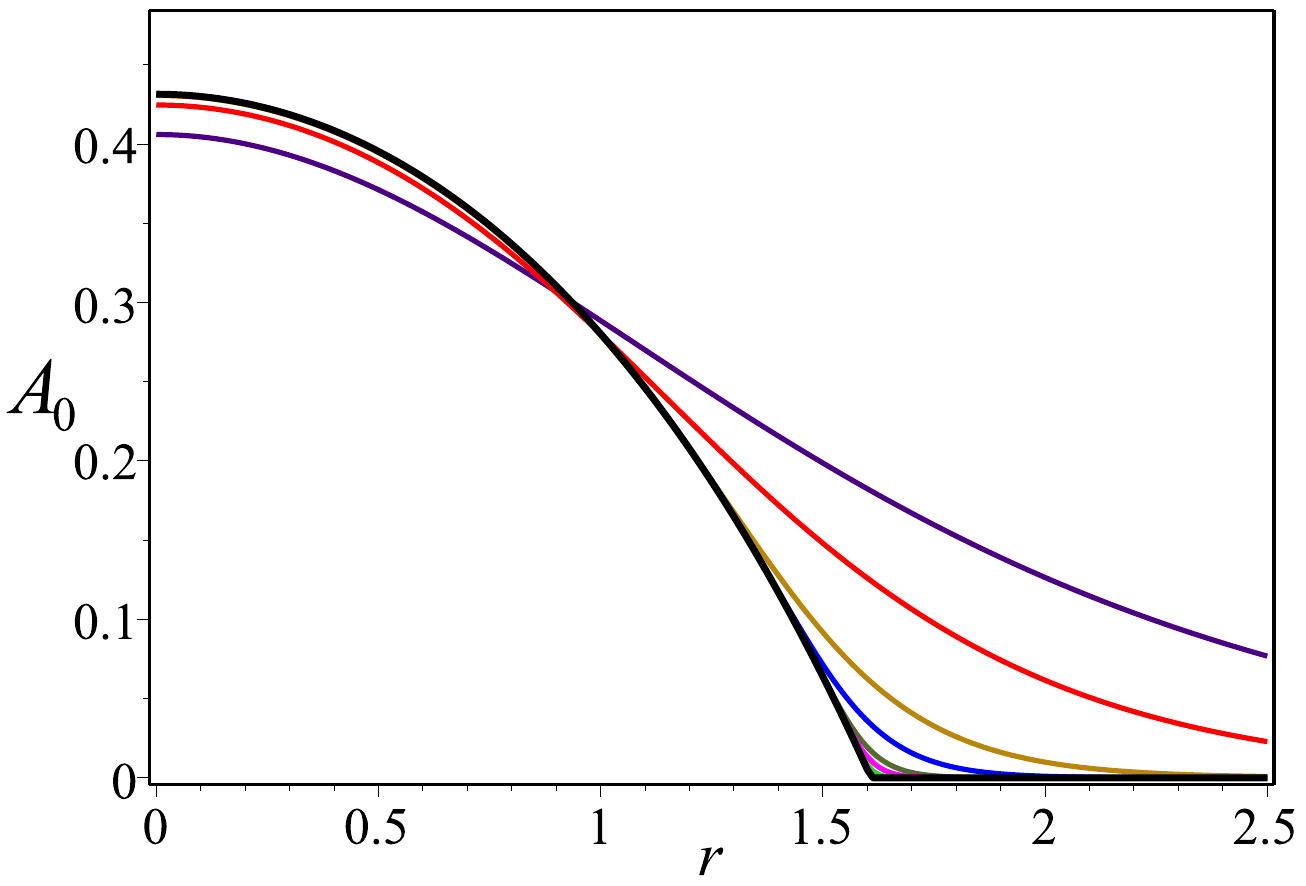} \centering%
\includegraphics[width=8.6cm]{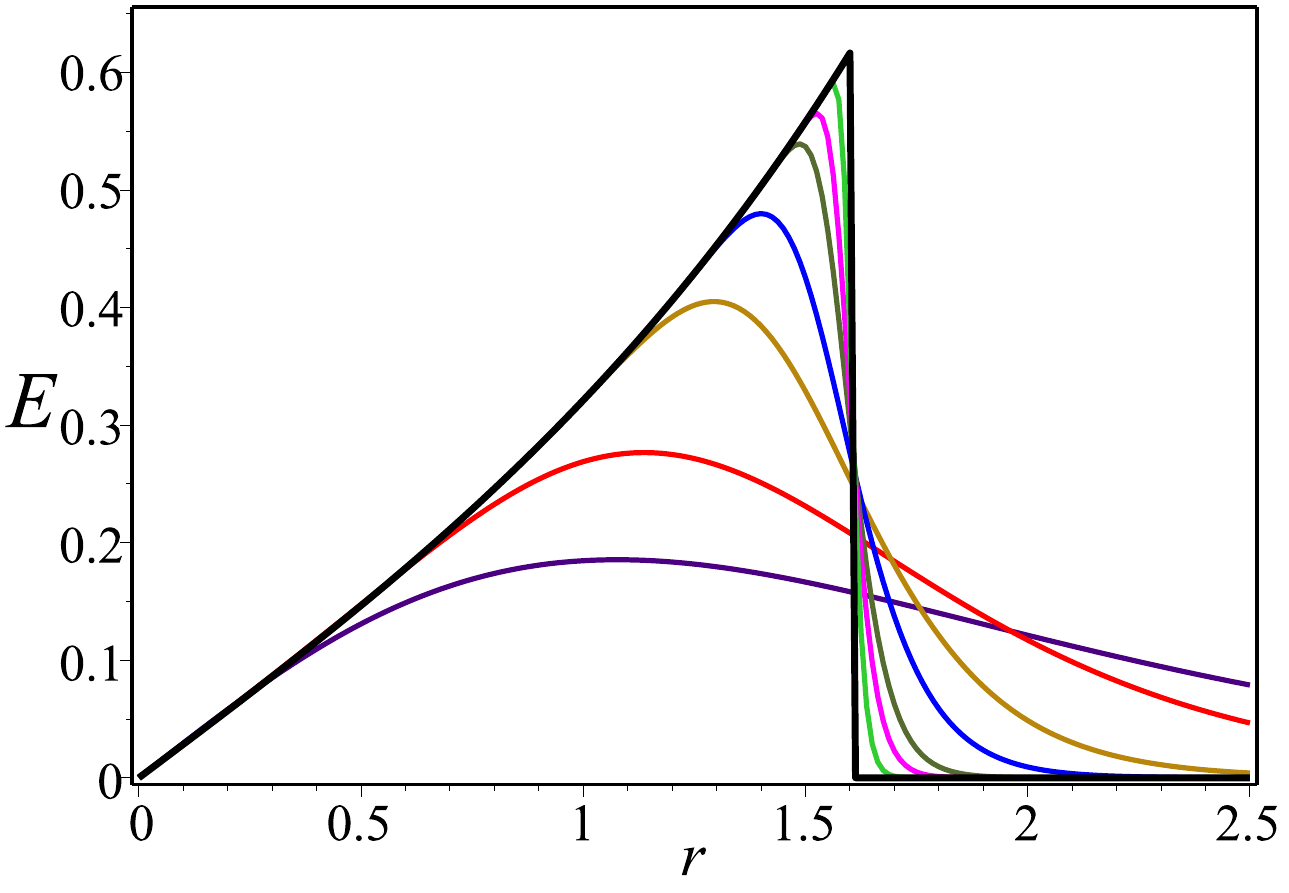}
\caption{The scalar
potential $A_0(r)$ (upper) and the electric field $E(r)$ (lower) coming from
generalized Maxwell-Chern-Simons-Higgs model (\protect\ref{Acg2}) with
$G(g)=1$ and $\protect\omega(g)=\protect\lambda g^{2\protect\lambda-2}$.
Observe that $\lambda=1$ (indigo lines) represents the usual MCSH model and
the true compacton solution is given by $\lambda=\infty$ (black lines).} %
\label{fig08}
\end{figure}

The compacton boundary conditions satisfied by the profiles $g(r)$, $a(r)$
and $A_{0}(r)$ are
\begin{eqnarray}
g(0) &=&0,\;\;a(0)=n,\;\;A_{0}^{\prime }(0)=0, \\[0.2cm]
g(r) &=&1,\;\;a(r)=0,\;\;A_{0}(r)=0, \;\;r_{c}\leq r<\infty. \quad
\end{eqnarray}%
The radial distance $r_{c}<\infty $ is the value where the profile $g(r)$
reaches the vacuum value, the gauge field profile $a(r)$ and scalar
potential $A_{0}(r)$ becomes null.

The system is solved analytically to be
\begin{eqnarray}
g^{(\infty )}(r) &=&\left(\frac{r}{r_{c}}\right)^n \exp \left[ \frac{e^{2}v^{2}}{\kappa ^{2}}%
\left( 1-\frac{I(0,\kappa r)}{I(0,\kappa r_{c})}\right) \right] \Theta
\left( r_{c}-r\right)  \notag \\[0.2cm]
&&+\Theta \left( r-r_{c}\right), \\[0.2cm]
a^{(\infty )}(r) &=&n\left( 1-\frac{rI(1,\kappa r)}{r_{c}I(1,\kappa r_{c})}\right)
\Theta \left( r_{c}-r\right) ,\\[0.2cm]
A^{(\infty )}_{0}(r) &=&\frac{ev^{2}}{\kappa }\left( -1+\frac{I(0,\kappa r)}{I(0,\kappa
r_{c})}\right) \Theta \left( r_{c}-r\right).
\end{eqnarray}%
The radial distance $r_{c}$ is computed from the equation%
\begin{equation}
I(0,\kappa r_{c})=r_{c}\frac{e^{2}v^{2}}{n\kappa }I(1,\kappa r_{c}),
\end{equation}%
where the function $I(\nu ,x)$ is the modified Bessel function of the first
kind and order $\nu $.

The magnetic field and BPS energy density of the Maxwell-Chern-Simons-Higgs compacton are
\begin{eqnarray}
B^{(\infty )}(r) &=&ev^{2}\frac{I(0,\kappa r)}{I(0,\kappa r_{c})}\Theta
(r_{c}-r), \\[0.2cm]
\varepsilon _{_{BPS}}^{(\infty )}(r) &=&e^{2}v^{4}\left( \frac{I(0,\kappa r)%
}{I(0,\kappa r_{c})}\right) ^{2}\Theta (r_{c}-r) \\[0.2cm]
& & +e^{2}v^{4}\left( \frac{I(1,\kappa r)}{I(0,\kappa r_{c})}\right) ^{2}
\Theta (r_{c}-r).  \notag
\end{eqnarray}

\begin{figure}[]
\centering\includegraphics[width=8.6cm]{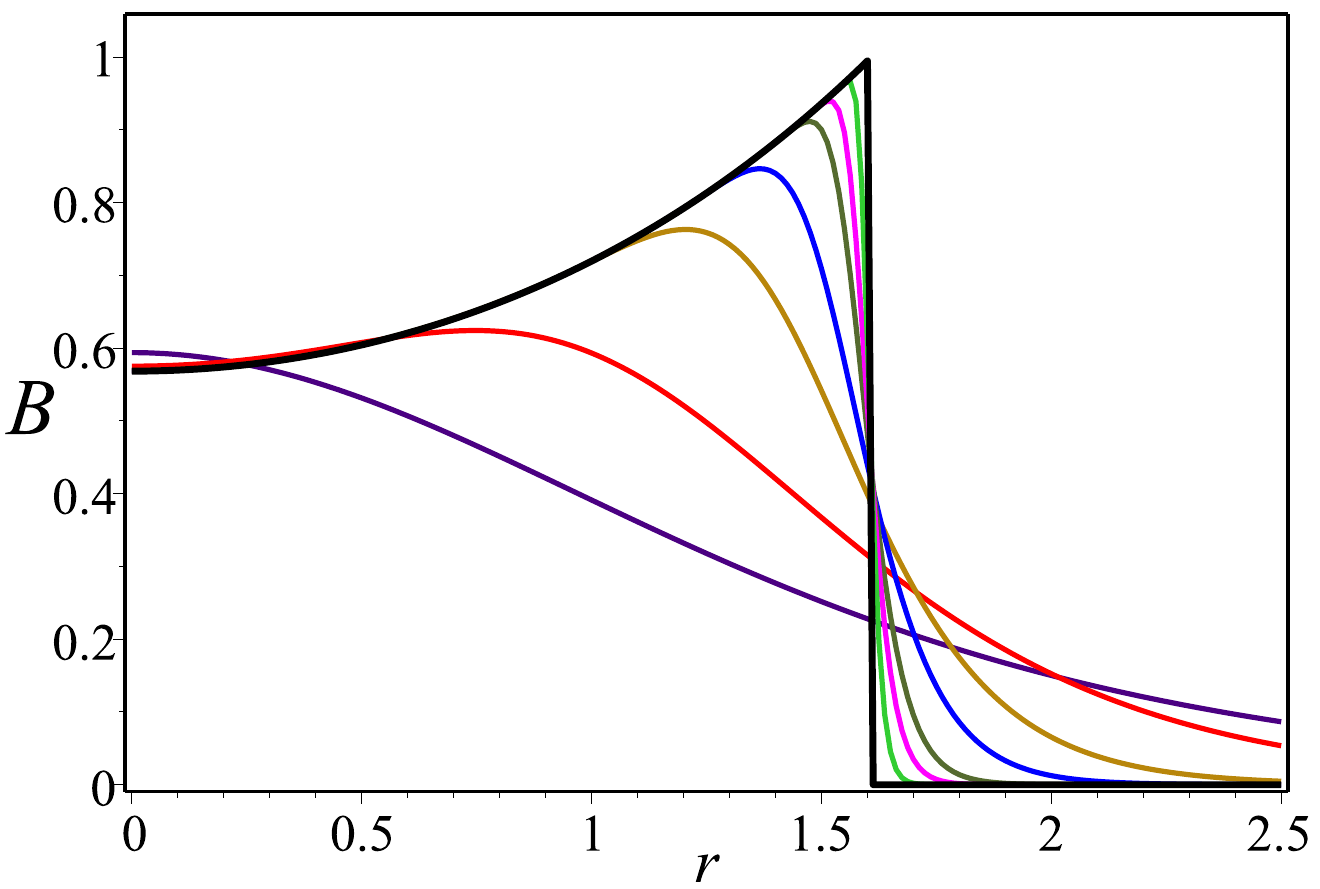} \centering%
\includegraphics[width=8.6cm]{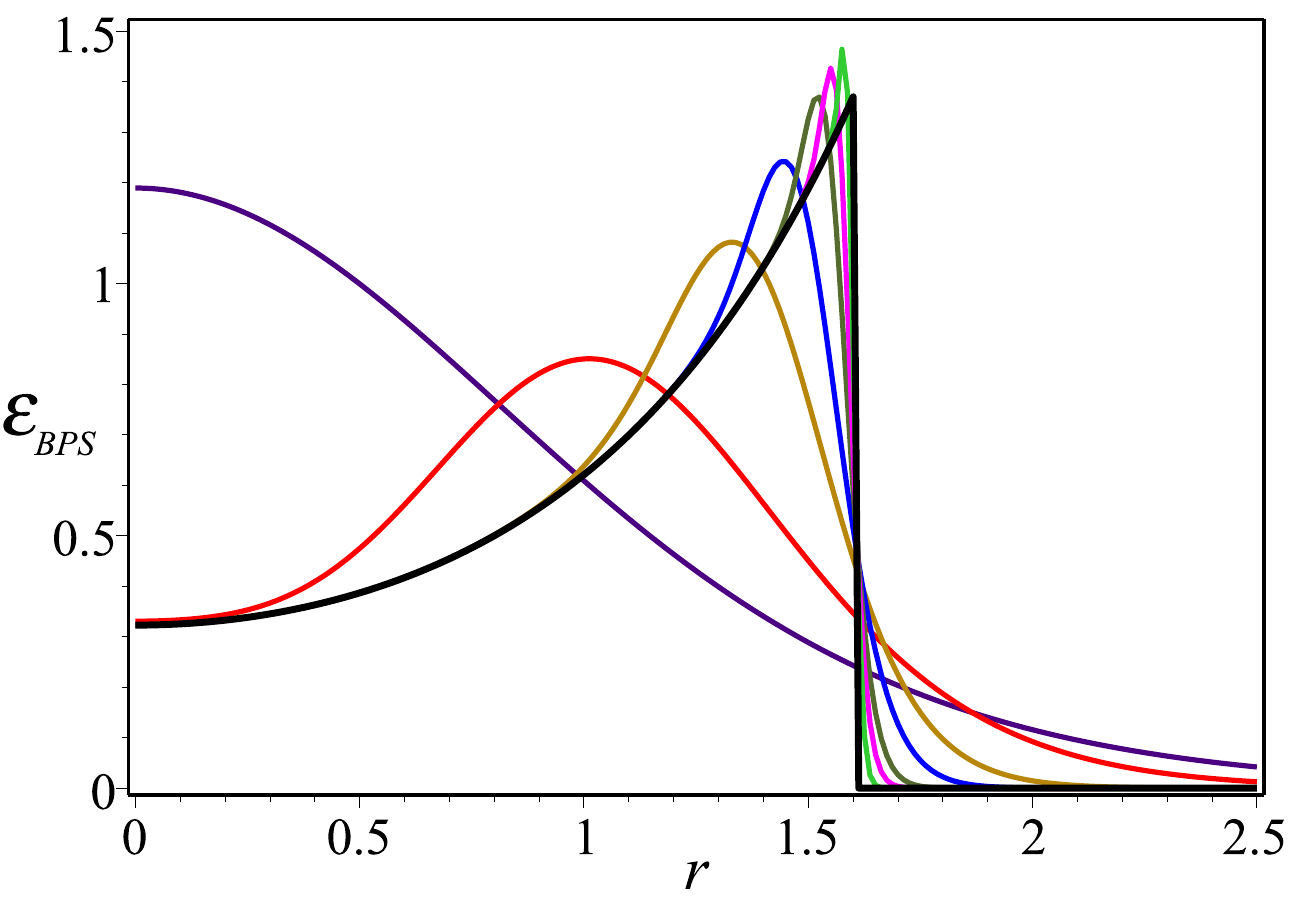}
\caption{The magnetic field $B(r)$ (upper) and the BPS energy density $\protect\varepsilon _{_{BPS}}(r)$ (lower) coming from generalized
Maxwell-Chern-Simons-Higgs model (\protect\ref{Acg2}) with $G(g)=1$ and $\protect\omega(g)=\protect\lambda g^{2\protect\lambda-2}$. Observe that $\lambda=1$ (indigo lines) represents the usual MCSH model and the true compacton solution is given by $\lambda=\infty$ (black lines).}
\label{fig09}
\end{figure}

In order to compute the numerical solutions we choose the upper signs in
equations (\ref{mcshb1}) and (\ref{mcshb2}), $e=1$, $v=1$, $\kappa =-1$ and
winding number $n=1$.  The profiles for the Higgs and gauge fields are
given in Fig. \ref{fig07},  the correspondent ones for the scalar potential and
for the electric field are depicted in Fig. \ref{fig08}. We can note again that
an effective compact topological defect it is formed for large values of
$\lambda$. This feature can be seen from the magnetic field and BPS energy
density profiles in Fig. \ref{fig09}. Alike in the previous studied models the
analytic MCSH compactons are formed for $\lambda=\infty$, they are represented by the solid black lines in Figs. \ref{fig07}, \ref{fig08} and \ref{fig09}.

\section{Remarks and conclusions}

We have found self-dual or BPS configurations in Abelian-Higgs generalized
models which given origin to new  effective compact and true compacton
configurations. Our goal was obtained by means of a consistent implementation
of the BPS formalism which besides to provide the self-dual or BPS equations
have also allowed to found the explicit form of the generalizing function
$\omega(|\phi|)$ (see Eq. (\ref{omegaa})) which is parameterized by the
positive parameter $\lambda$. Such a parameter determine explicitly new
families of self-dual potentials for every model and consequently characterize
their self-dual configurations. We draw attention to the importance to
obtain self-dual effective compact and analytic true compacton configurations in Abelian Higgs. This models enhance the space of self-dual solutions which probably will imply in interesting applications in physics and mathematics, for example, the construction of the respective supersymmetric extensions \cite{SUSY}.

For every model we have studied the vortex solutions arising from the
respective self-dual equations. The numerical analysis have shown that for
sufficiently large values of $\lambda $ the profiles (of the Higgs field, gauge
field, magnetic field, BPS energy density) are very alike with the ones of an
effective compacton solution but still preserve a tail in their asymptotic
decay. For every model, we have also analyzed the limit $\lambda\rightarrow
\infty$ for arbitrary winding number $n$. Our analysis have shown that for
$\lambda=\infty$ arises analytical compacton structures in all models (see black line profiles in all figures along the manuscript).

Finally, we are considering the interesting challenge of looking for effective compact structures in gauge field models which engender monopoles or skyrmions, for example. Advances in this direction will be reported elsewhere.

\section*{Competing Interests}

The authors declare that there is no conflict of interests regarding the
publication of this paper.

\begin{acknowledgments}
RC and GL thank to CNPq, CAPES and FAPEMA (Brazilian agencies) by financial
support. L.S. is supported by CONICET.
\end{acknowledgments}

\end{document}